\documentclass[11pt]{article}
\pdfoutput=1

\usepackage{cancel,slashed}
\usepackage{bbm}
\usepackage{bm}
\usepackage{amsmath,amsfonts,amssymb,mathtools,nicefrac,nccmath,cases}
\usepackage{graphicx}
\usepackage{cite}
\usepackage{skak}

\usepackage{dsfont}
\usepackage{latexsym}
\usepackage{color}
\usepackage[hyperfootnotes=false,linktocpage]{hyperref}
\usepackage{transparent}
\usepackage[hang,flushmargin]{footmisc}
\usepackage{blkarray}
\usepackage{multirow}
\usepackage{empheq}
\usepackage{comment}
\usepackage{wasysym}
\usepackage{tikzsymbols}

\usepackage{caption}
\usepackage{subcaption}
\usepackage[normalem]{ulem}
\usepackage{braket}

\newcommand{\bmat}{\left(\begin{array}}
\newcommand{\emat}{\end{array}\right)}

\def\a {\alpha}
\def\b {\beta}

\def\1{{\bf 1}}
\def\2{{\bf 2}}
\def\3{{\bf 3}}
\def\4{{\bf 4}}
\def\6{{\bf 6}}

\def\targ#1#2{\genfrac{[}{]}{0pt}{}{#1}{#2}}
\def\targ2#1#2{\genfrac{}{}{0pt}{}{#1}{#2}}

\definecolor{mygr}{rgb}{0,0.6,0}
\definecolor{mygrey}{rgb}{0,0.1,0.2}
\definecolor{myblue}{rgb}{0,0.5,0.9}
\definecolor{myblue2}{rgb}{0,0.5,0.5}
\definecolor{myblue3}{rgb}{0,0.7,0.9}
\definecolor{myblue4}{rgb}{0,0.6,0.6}
\definecolor{myorange}{rgb}{1,0.5,0}
\definecolor{mypurple}{rgb}{0.6,0,1}
\definecolor{mygolden}{rgb}{1,0.8,0.2}
\definecolor{mycyan}{rgb}{0,1,1}
\definecolor{mymagenta}{rgb}{1,0,1}
\definecolor{mykiwi}{rgb}{0.8,1,0.5}
\definecolor{mybrown}{cmyk}{0.14, 0.42, 0.56, 0.2}
\definecolor{myturq}{cmyk}{0.99, 0, 0.2, 0.4}
\definecolor{myaubergine2}{cmyk}{0.4, 0.5, 0, 0.1}
\definecolor{myaubergine}{cmyk}{0.6,0.85,0,0}
\definecolor{CycleGreen}{cmyk}{0.52,0,1,0}
\definecolor{CycleBrown}{cmyk}{0, 0.4, 0.9, 0.2}

\DeclareFontFamily{U}{rcjhbltx}{}
\DeclareFontShape{U}{rcjhbltx}{m}{n}{<->rcjhbltx}{}
\DeclareSymbolFont{hebrewletters}{U}{rcjhbltx}{m}{n}

\DeclareMathSymbol{\lamed}{\mathord}{hebrewletters}{108}
\DeclareMathSymbol{\mem}{\mathord}{hebrewletters}{109}
\DeclareMathSymbol{\ayin}{\mathord}{hebrewletters}{96}
\DeclareMathSymbol{\tsadi}{\mathord}{hebrewletters}{118}
\DeclareMathSymbol{\qof}{\mathord}{hebrewletters}{113}
\DeclareMathSymbol{\resh}{\mathord}{hebrewletters}{114}
\DeclareMathSymbol{\pe}{\mathord}{hebrewletters}{112}
\DeclareMathSymbol{\pesofit}{\mathord}{hebrewletters}{80}
\DeclareMathSymbol{\samekh}{\mathord}{hebrewletters}{115}
\DeclareMathSymbol{\tav}{\mathord}{hebrewletters}{116}
\DeclareMathSymbol{\vav}{\mathord}{hebrewletters}{119}
\DeclareMathSymbol{\het}{\mathord}{hebrewletters}{120}
\DeclareMathSymbol{\yod}{\mathord}{hebrewletters}{121}
\DeclareMathSymbol{\zayin}{\mathord}{hebrewletters}{122}
\DeclareMathSymbol{\alephdot}{\mathord}{hebrewletters}{128}
\DeclareMathSymbol{\tsadisofit}{\mathord}{hebrewletters}{90}
\DeclareMathSymbol{\shin}{\mathord}{hebrewletters}{152}

\newtheorem{conjecture}{Conjecture}

\def\CN {{\cal N}}

\def\be{\begin{equation}}
\def\ee{\end{equation}}
\def\bea{\begin{eqnarray}}
\def\eea{\end{eqnarray}}
\def\bes{\begin{subequations}}
\def\ees{\end{subequations}}

\def\eps{{\epsilon}}
\def\oh{\frac{1}{2}}

\def\re{\mbox{Re}\, }
\def\im{\mbox{Im}\, }

\def\om{\omega}
\def\Om{\Omega}
\usepackage{multicol}
\usepackage{float}
\usepackage{caption}
\def\p {{\partial}}

\def\g {{\gamma}}

\def\th {{\theta}}

\newcommand{\cF}{\mathcal{F}}

\newcommand{\IR}{\mathbb{R}}
\newcommand{\IZ}{\mathbb{Z}}

\newenvironment{eqn*}{\begin{equation*}\begin{aligned}}{\end{aligned}\end{equation*}\noindent}

\makeatletter
\newsavebox\myboxA
\newsavebox\myboxB
\newlength\mylenA

\newcommand*\xoverline[2][0.75]{%
\sbox{\myboxA}{$\m@th#2$}%
\setbox\myboxB\null% Phantom box
\ht\myboxB=\ht\myboxA%
\dp\myboxB=\dp\myboxA%
\wd\myboxB=#1\wd\myboxA% Scale phantom
\sbox\myboxB{$\m@th\overline{\copy\myboxB}$}%  Overlined phantom
\setlength\mylenA{\the\wd\myboxA}%   calc width diff
\addtolength\mylenA{-\the\wd\myboxB}%
\ifdim\wd\myboxB<\wd\myboxA%
   \rlap{\hskip 0.5\mylenA\usebox\myboxB}{\usebox\myboxA}%
\else
    \hskip -0.5\mylenA\rlap{\usebox\myboxA}{\hskip 0.5\mylenA\usebox\myboxB}%
\fi}
\makeatother

\begin{document}
\pagestyle{plain}

%----------------------------------------------------------------------%
%  numbering equations with section number
%----------------------------------------------------------------------%
\makeatletter
\@addtoreset{equation}{section}
\makeatother
\renewcommand{\theequation}{\thesection.\arabic{equation}}
%----------------------------------------------------------------------%
%  title page
%----------------------------------------------------------------------%

\pagestyle{empty}
%\vspace*{1.0in}
\rightline{IFT-UAM/CSIC-23-77}
%\rightline{\tt hep-th/yymmnnn}
\vspace{0.5cm}
\begin{center}
\Huge{{Torsion in cohomology \\ and dimensional reduction} 
\\[10mm]}
\normalsize{Gonzalo F. Casas,$^{1}$ Fernando Marchesano\,$^{1}$  and Matteo Zatti}\,$^{1,2}$\\[12mm]
\small{
${}^{1}$ Instituto de F\'{\i}sica Te\'orica UAM-CSIC, c/ Nicol\'as Cabrera 13-15, 28049 Madrid, Spain \\[2mm] 
${}^{2}$ Departamento de F\'{\i}sica Te\'orica, Universidad Aut\'onoma de Madrid, 28049 Madrid, Spain
\\[10mm]} 
\small{\bf Abstract} \\[5mm]
\end{center}
\begin{center}
\begin{minipage}[h]{15.0cm} 

Conventional wisdom dictates that $\IZ_N$ factors in the integral cohomology group $H^p(X_n, \IZ)$ of a compact manifold $X_n$ cannot be computed via smooth $p$-forms. We revisit this lore in light of the dimensional reduction of  string theory on $X_n$, endowed with a $G$-structure metric that leads to a supersymmetric EFT. If massive $p$-form eigenmodes of the Laplacian enter the EFT, then torsion cycles coupling to them will have a non-trivial smeared delta form, that is an EFT long-wavelength description of $p$-form currents of the $(n-p)$-cycles of $X_n$. We conjecture that, whenever torsion cycles are calibrated, their linking number can be computed via their smeared delta forms. From the EFT viewpoint, a torsion  factor in cohomology corresponds to a $\IZ_N$ gauge symmetry realised by a St\"uckelberg-like action, and calibrated torsion cycles to BPS objects that source the massive fields involved in it.

\end{minipage}
\end{center}
\newpage
%----------------------------------------------------------------------%
%  Resetting of counters
%----------------------------------------------------------------------%
\setcounter{page}{1}
\pagestyle{plain}
\renewcommand{\thefootnote}{\arabic{footnote}}
\setcounter{footnote}{0}
%----------------------------------------------------------------------%
%  Paper begins
%----------------------------------------------------------------------%

%\end{document}

\tableofcontents

%%%%%%%%%%%%%%%%%%%
%%%%%%%%%%%%%%%%%%%
%\newpage

\section{Introduction}
\label{s:intro}

String theory compactifications provide a remarkable connection between the geometry of extra dimensions and the physics of Effective Field Theories (EFTs) \cite{Becker:2006dvp,Ibanez:2012zz,Blumenhagen:2013fgp,Baumann:2014nda,Tomasiello:2022dwe}. An early lesson that one obtains upon exploring this link is that the more an EFT quantity is protected against quantum corrections, the simpler is its description in geometric terms. Typical examples arise in the context of supergravity and supersymmetric gauge theories, where protection mechanisms involve both gauge invariance and renormalisation effects constrained by supersymmetry. 

In several instances, discrete EFT data protected by gauge invariance are described in terms of the topology of the compact manifold $X_n$, while quantities protected by supersymmetry enjoy a simple description in terms of differential and/or algebraic geometry. A clear example of the second is BPS states or extended objects of the EFT, which can be obtained from, e.g., D-branes in type II compactifications at weak coupling. In that case, the BPSness condition requires that the D-brane extra dimensions wrap a $p$-cycle of $X_n$ that is calibrated, in the sense of \cite{Harvey:1982xk}. This condition not only has a neat differential geometric description for compactification manifolds $X_n$ with special holonomy, but it can be generalised whenever $X_n$ has a $G$-structure metric and a flux background that leads to a supersymmetric EFT \cite{Martucci:2005ht,Koerber:2005qi,Koerber:2010bx}. The central charge of the BPS object at tree-level is then determined by the integral over the $p$-cycle of the suitable $p$-form calibration, or generalisations that allow us to calibrate D-brane bound states. This picture also applies to space-time filling D-branes that are part of the background in type II orientifold compactifications, as well as to Euclidean D-branes that play the role of BPS instantons.  

An example of discrete EFT data with a topological higher-dimensional origin is the presence of discrete gauge symmetries. In the Abelian case, a $\IZ_N$ gauge symmetry of a $d$-dimensional EFT can be described by a Lagrangian coupling of the form \cite{Banks:2010zn}
\be
N  \, B_{d-2} \wedge F_2 ,
\label{BF}
\ee
where $B_{d-2}$ is a $(d-2)$-form of the EFT  dual to an axion $C_0$, and $F_2=dA_1$ is the field strength of the $U(1)$ boson gauged by $C_0$  \`a la St\"uckelberg. Finally, $N \in \IZ$ is the quantity that is described in terms of the topology of $X_n$. For instance, in type II orientifold compactifications, couplings of this form are specified by the homology classes of the $p$-cycles wrapped by space-time filling D-branes, which in turn determine the discrete gauge symmetries acting on the open string sector of the theory \cite{Berasaluce-Gonzalez:2011gos}. This case is particularly interesting because the discrete symmetry acts on the massless chiral spectrum of the EFT. However, it has the feature that the axion and gauge boson masses induced by \eqref{BF} are usually of the order of the string scale. This implies that the St\"uckelberg terms that complete  \eqref{BF} are not part of the EFT Lagrangian. 

A different setup where the coupling \eqref{BF} is realised is by threading the compact manifold $X_n$ with quantised background fluxes \cite{Berasaluce-Gonzalez:2012awn}. In this case, the coupling $N$ is determined by the flux quanta, or equivalently by an integral cohomology class in $X_n$. Here the interplay with the EFT cutoff is reversed with respect to the previous one. The St\"uckelberg-induced masses for axions and gauge bosons can lie below the EFT cutoff, but now the resulting discrete gauge symmetry acts on strings and particles that typically do not correspond to light states of the EFT. 

In this work, we are interested in yet another realisation of discrete gauge symmetries, namely those that arise from torsion factors in the integral cohomology groups $H^p(X_n, \IZ)$. That such $\IZ_N$ factors correspond to $\IZ_N$ gauge symmetries can be seen in the AdS/CFT context by following the reasoning in \cite{Gukov:1998kn,Witten:1998wy},  applied to type II orientifold compactifications in \cite{Camara:2011jg}, and with subsequent work in similar setups in \cite{Grimm:2011tb,Berasaluce-Gonzalez:2012abm,Berasaluce-Gonzalez:2012awn,Mayrhofer:2014laa,Grimm:2015ona,Braun:2017oak}. As stressed in \cite{Berasaluce-Gonzalez:2012awn}, the realisation of discrete gauge symmetries via torsion in cohomology is related to the setting with background fluxes by dualities such as mirror symmetry. This implies that the same EFT features should be realised, namely: {\it i)} St\"uckelberg couplings that are part of the EFT Lagrangian and {\it ii)}  charged objects that lie above the EFT cut-off. Indeed, as discussed in \cite{Camara:2011jg} such charged objects are given by D-branes wrapping torsion cycles of $X_n$, which from the EFT perspective look like particles and $(d-3)$-branes coupling to $A_1$ and $B_{d-2}$, respectively. 

From this simple observation, an apparent puzzle follows. If in this case \eqref{BF} and its St\"uckelberg  completion appears in the lower-dimensional EFT, is because torsion in cohomology is detected by the standard procedure of $p$-form dimensional reduction. This is rather counter-intuitive, in the sense that torsion cohomology groups are trivial in de Rham cohomology, or in other words their elements can only be represented by exact $p$-forms. Since the EFT data captured upon dimensional reduction typically involves integrals of $p$-forms over $p$-cycles, it is a priori not clear how torsion cohomology factors can translate into a St\"uckelberg  Lagrangian term in the EFT. This naive picture agrees with the standard lore that torsion in cohomology cannot be detected via smooth $p$-forms, and that one should resort to more advanced geometric techniques, like the computation of spectral sequences  \cite{Bott1982DifferentialFI} or to differential cohomology \cite{Apruzzi:2021nmk}.

This paper addresses this puzzle and proposes a prescription to capture torsion in cohomology via the standard procedure of dimensional reduction. The basic idea is to use {\em smeared delta forms} to construct the integral basis in which ten-dimensional fields are expanded in order to perform the reduction. Here a delta $p$-form stands for the $p$-current $\delta_p(\Pi_{n-p})$ with legs transverse to an $(n-p)$-cycle $\Pi_{n-p} \subset X_n$, while its smeared version $\delta_p^{\rm sm}(\Pi_{n-p})$ corresponds to the projection into the light eigen-$p$-forms of the Laplacian. If one projects $\delta_p(\Pi_{n-p})$ to the zero-mode sector of the spectrum one simply obtains a harmonic $p$-form which is the de Rham Poincar\'e dual of $\Pi_{n-p}$, and torsion cycles are projected out. If however, one includes in the projection those non-vanishing eigenmodes that correspond to massive $p$-form fields entering the EFT, then torsion cycles can have a non-trivial smeared delta form, and translate into quantities of the EFT Lagrangian. 

More precisely, we propose that one should consider smeared delta forms of calibrated cycles in order to build the basis for the dimensional reduction. The physical intuition behind this proposal is that D-branes wrapping calibrated cycles correspond to BPS objects with a controlled backreaction, that one can use together with the picture developed in \cite{Goldberger:2001tn,Michel:2014lva,Polchinski:2015bea} to see their smeared delta function as an EFT long-wavelength description of the corresponding object.  This can then be used to extract information from the EFT, as done in \cite{Lanza:2020qmt,Lanza:2021udy,Lanza:2022zyg} in the context of 4d $\CN=1$ compactifications. In particular, D-branes wrapping calibrated torsion cycles can be seen as BPS operators that gather information on the massive sector of the EFT Lagrangian, like the kinetic terms of the fields that appear in \eqref{BF}. As a direct consequence of our proposal, the linking number between two calibrated torsion cycles can be computed using EFT data, or equivalently by defining a smeared version of the torsion linking number, as summarised in Conjecture \ref{conj:BPS}. 

The notion of calibrated torsion cycle or BPS operator with a $\IZ_N$ charge may seem puzzling. From a geometric viewpoint, calibrations in special 
holonomy manifolds are closed $p$-forms, and therefore they can never calibrate a torsion $p$-cycle. This obstruction is however absent in the more general set of manifolds with $G$-structure metrics, since there the exterior derivative of a calibration does not need to vanish, and one can indeed construct explicit examples with torsion $p$-cycles that are calibrated. From a physics viewpoint, due to the no-force condition between mutually BPS objects, one should always be able to stack an arbitrary number of them on top of each other without any binding energy. This fits naturally with a $\IZ$-valued charge, but not with a $\IZ_N$-valued one. To address this issue we construct examples of BPS objects with $\IZ_N$ charge, in the context of domain-wall solutions of type II string compactified on half-flat manifolds \cite{Gurrieri:2002wz}. We find that the process that annihilates $N$ BPS D-branes wrapped on a torsion cycle is indeed possible topologically, but not energetically favoured. As a result, it is energetically stable to stack an arbitrary number of such objects with $\IZ_N$ charge, as implied by the BPS condition. 

Expressing a delta form as a sum of eigenforms of the Laplacian is in general involved, as it requires knowledge of the massive $p$-form spectrum of a manifold. This difficulty is however less severe for three-dimensional  manifolds with isometries, a fact that we exploit to perform an explicit computation of a torsion linking number and its smeared version in twisted tori, in order to verify Conjecture \ref{conj:BPS}. While it seems challenging to extend such a computation to more general setups, one can provide physical evidence that our proposal should also be valid in SU(3)-structure manifolds. Indeed, using smeared delta forms of calibrated cycles as a basis for dimensional reduction fits perfectly with the framework developed in \cite{DAuria:2004kwe,Grana:2005ny,Kashani-Poor:2006ofe} to describe 4d $\CN=2$ gauged supergravities as EFTs of type II string compactifications and, in fact, one may argue that it is necessary for the consistency of the approach. Similar considerations can be drawn in the context of 4d $\CN=1$ type II orientifold vacua, where the BPS torsion objects are given by membranes ending on strings, and by space-time filling branes ending on membranes. 

In most of our examples it seems that an extension of Conjecture \ref{conj:BPS} is required. In such a generalisation, the torsion linking number can be computed not only when elements of ${\rm Tor}H_{n-p}(X_n, \IZ)$ contain calibrated representatives, but also when they can be expressed as linear combinations of elements of $H_{n-p}(X_n, \IZ)$, all of them with calibrated representatives. This extension could in principle be applied to compute torsion factors in the cohomology of Calabi--Yau manifolds, whenever they contain light eigenforms of the Laplacian other than harmonic forms. One may even speculate that our approach could be useful to compute torsion linking numbers even in cases where torsion cycles cannot be related to calibrated submanifolds, by providing an estimate of the associated error in the linking number computation. In any event, our findings support that one may compute certain torsion topological invariants in terms of smeared or EFT data such as masses and kinetic terms, extending the dictionary between geometry and physics to the more subtle and unexplored sector that is torsion in cohomology. 

The rest of the paper is organised as follows. In section \ref{s:proposal} we describe what is our proposal to compute the torsion linking numbers of a manifold via smeared delta forms, as well as an extension of such proposal. In section \ref{s:dimred} we motivate the proposal from a physics viewpoint, by interpreting torsion calibrated cycles as BPS objects of the EFT with a non-trivial backreaction. In section \ref{s:simple} we analyse our proposal in the context of domain-wall solutions of 4d $\CN=2$ EFTs obtained from compactifications of type IIA string theory on half-flat manifolds. The simplest examples of such manifolds are based on twisted three-tori, for which our conjecture can be verified explicitly using the techniques of section \ref{s:direct}. Section \ref{s:general} tests our proposal in the context of general SU(3)-structure compactifications of type IIA string theory, finding agreement with previous analysis in the literature and giving a more precise prescription to perform the dimensional reduction in this context. Section \ref{s:N=1} extends our general strategy to 4d $\CN=1$  type II orientifold vacua, and section \ref{s:nonBPS} contains some speculative remarks on how to perform a further extension to the case where no EFT BPS objects are available to detect torsion. We finally draw our conclusions in section \ref{s:conclu}. 

Several technical details have been relegated to the appendices. Appendix \ref{ap:NS5DW} analyses a mirror dual setup to that of section \ref{s:simple} from a microscopic viewpoint, in order to classify the relevant set of BPS D-branes in both backgrounds. Appendix \ref{ap:spectra} analyses the massive $p$-form spectrum for the case of the twisted three-torus, as a necessary step to perform the direct computation of the torsion linking number of section \ref{s:direct}.

%%%%%%%%%%%%%%%%%%%
%%%%%%%%%%%%%%%%%%%

\section{The proposal}
\label{s:proposal}

Let us consider a compact manifold $X_n$ of real dimension $n$, and a submanifold $\Pi_p \subset X_n$ which is a $p$-cycle. We can define a bump-delta $(n-p)$-current or distributional form $\delta (\Pi_p)$, such that
\be
\int_{X_n} \omega_p \wedge \delta (\Pi_p) = \int_{\Pi_p} \omega_p \, , 
\label{deltadef}
\ee
for any smooth $p$-form $\omega_p \in \Omega^p (X_n)$. If $X_n$ is endowed with a smooth metric $ds^2_{X^n}$ measured in string units $\ell_s = 2\pi \sqrt{\a'}$, one can solve the eigenvalue problem for $(n-p)$-forms
\be
\Delta b_{n-p}^i = \lambda_i^2 b_{n-p}^i \, , 
\ee
where $\Delta = d^\dag d + d d^\dag$ is the Laplace-de Rham operator, $\{ b_{n-p}^i\}_i$ is an orthonormal basis of eigenforms with respect to the Hodge product, and $\{\lambda_i^2\}_i$ the corresponding set of non-negative, dimensionless eigenvalues. Then one can expand the bump-delta $(n-p)$-form on such a basis
\be
\delta (\Pi_p) = \sum_i c_i  \, b^i_{n-p}\, , \qquad c_i = \int_{\Pi_p} *  b^i_{n-p}\, ,
\label{deltaexp}
\ee
and from here define a {\em smeared} version of the delta-form, by keeping only those terms in the expansion that satisfy $\lambda_i < \lambda_{\rm max}$, for some choice of $\lambda_{\rm max}$. This  defines a smooth bump $(n-p)$-form localised within a tubular neighbourhood of radius $\ell_s/\lambda_{\rm max}$ around $\Pi_p$. Such a $(n-p)$-form can be identified with the Thom class of the normal bundle of $\Pi_p$, which is known to lie in the de Rham Poincar\'e dual to $[\Pi_p] \in H_p(X_n)$ \cite{Bott1982DifferentialFI}. Indeed, notice that all elements $b^{i}_{n-p}$ of the expansion \eqref{deltaexp} must be exact $(n-p)$-forms except those with vanishing eigenvalue, which must correspond to the harmonic representative of the Poincar\'e dual to $[\Pi_p]$. It follows that if $[\Pi_p]$ lies in a torsion class of $H_p(X_p, \IZ)$ then $\delta(\Pi_p)$ must be a sum of exact $(n-p)$-forms.

In string theory compactifications there is a natural choice of metric for $X_n$ that comes from solving the 10d supergravity equations of motion, as well as a natural choice of $\lambda_{\rm max}$ that one identifies with the compactification scale $m_{\rm KK}$. One can define $\ell_s m_{\rm KK}$ to be the typical spacing between positive eigenvalues $\lambda_i$, oftentimes estimated by the average radius Vol$(X_n)^{1/n}$. Physically, we understand $m_{\rm KK}$ as the energy scale below which we recover a $D$-dimensional EFT description, with $D=10-n$, that describes all eigenmodes with $\lambda_i \ll \ell_s m_{\rm KK}$  as $D$-dimensional fields. The standard practice in the string literature is to either assume that only harmonic modes satisfy the requirement $\lambda_i \ll \ell_s m_{\rm KK}$, such that the procedure of dimensional reduction simply projects the spectrum of $p$-forms to the harmonic sector.\footnote{Alternatively, one may set the EFT cut-off $\Lambda_{\rm EFT}$ below any non-vanishing mode.} However, it has been shown that in certain compactification regimes, and in particular in six-dimensional manifolds with SU(3)-structure \cite{Gray1980TheSC,chiossi2002intrinsic,Hull:1986iu,Strominger:1986uh,LopesCardoso:2002vpf,Gauntlett:2003cy,Grana:2005jc,Koerber:2010bx}, one has a non-vanishing $p$-form eigenvalues well below the compactification scale. This will be the case of interest in this work, and henceforth our definition of smeared delta-form will correspond to the following:
\be
\delta^{\rm sm} (\Pi_p) = \sum_{\lambda_i \ll \ell_s m_{\rm KK}} c_i  \, b^{i}_{n-p}\, .
\label{deltasm}
\ee

Note that if $[\Pi_p] \in {\rm Tor} H_p (X_p, \IZ)$ then \eqref{deltasm} may contain no terms at all and, if it does, it will be a sum of exact $(n-p)$-forms. This reflects the difficulties in obtaining information from the torsion (co)homology classes from the viewpoint of the lower dimensional EFT, as integrals of \eqref{deltasm} over any $(n-p)$-cycle of $X_n$  simply vanish. There is however a well-defined topological invariant  for torsion homology classes, which is the torsion linking number. Given the torsion classes $[\Pi_p] \in {\rm Tor} H_p(X_n, \IZ)$ and $[\Pi_{n-p-1}] \in {\rm Tor} H_{n-p-1}(X_n, \IZ)$, one can define their linking number in terms of the bump delta-forms of two of their representatives as \cite{Camara:2011jg}  
\be
L( \Pi_{n-p-1},\Pi_p) = \int_{X_n}  d^{-1}  \delta(\Pi_{n-p-1}) \wedge \delta(\Pi_p)  \quad \mod \ 1 \, .
\ee
Following \cite{Horowitz:1989km}, we can rewrite this quantity as follows. Notice that $\{\lambda_i ^{-1}d*b^i_{n-p}\}_i$ is an orthonormal basis of exact $(p+1)$-forms, so one can perform the expansion 
\be
\delta(\Pi_{n-p-1}) = \sum_i \frac{e_i}{\lambda_i} d*b^i_{n-p} \, ,  \qquad e_i = \frac{(-1)^{n(n-p)}}{\lambda_i}\int_{\Pi_{n-p-1}}d^\dag  b^i_{n-p}\, ,
\label{deltaexp2}
\ee
from where one obtains
\be
L( \Pi_{n-p-1},\Pi_p) = \sum_i \frac{c_ie_i}{\lambda_i}   \quad \mod \ 1 \, .
\label{L}
\ee
One can now define a {\em smeared linking number}. From $[\Delta, d] = [\Delta, *] = 0$ it follows that $d* b_{n-p}^i$ has the same eigenvalue as $b_{n-p}^i$, and so the smeared version of \eqref{deltaexp2} corresponds to the same truncation as in \eqref{deltaexp}. Thus, it is natural to define the smeared analogue of \eqref{L} as
\be
L^{\rm sm}( \Pi_{n-p-1},\Pi_p) = \sum_{\lambda_i \ll \ell_s m_{\rm KK}} \frac{c_ie_i}{\lambda_i}   \quad \mod \ 1 \, .
\label{Lsm}
\ee

On the one hand, this quantity is not a topological invariant of $X_n$. Unlike for \eqref{L}, there is no reason for it to remain invariant under a continuous deformation of either of the representatives $\Pi_p$ or $ \Pi_{n-p-1}$. On the other hand, as we argue in section \ref{s:dimred}, whenever \eqref{Lsm} is non-vanishing the massive sector of the $D$-dimensional EFT  knows about the value of \eqref{L}, so there must be some way in which one can find about this topological invariant in terms of smeared data.

In the following, we  propose a solution to this conundrum, namely that one needs to focus on certain minimal-volume representatives within the torsion homology class. More precisely, we consider manifolds $X_n$ that contain calibration $p$-forms, and torsion $p$-cycles that are calibrated by them. Calibration $p$-forms are standard objects in manifolds endowed with metrics of special holonomy \cite{Harvey:1982xk}, and from the viewpoint of string theory, they are present whenever $X_n$ leads to a $D$-dimensional supersymmetric EFT \cite{Martucci:2005ht,Koerber:2005qi,Koerber:2010bx}. In the case of special holonomy metrics calibrations are closed $p$-forms, and therefore torsion $p$-cycles cannot be calibrated. This is however different for manifolds with $G$-structure metrics, where several examples of calibrated torsion cycles exist. An illustrative case for our discussion in the following sections  will be the case of six-dimensional manifolds with SU(3)-structure, whose metric is specified by the pair of calibrations $(J, \Omega)$, which can respectively calibrate two- and three-cycles that are torsion or even trivial in homology. In terms of this language, our proposal can be expressed as follows:

\begin{conjecture}

A non-trivial smeared linking number between two calibrated torsion cycles equals their actual linking number. 

\label{conj:BPS}
\end{conjecture}

From a physics viewpoint, D-branes wrapping calibrated cycles correspond to BPS objects of different dimensions in the lower-dimensional supersymmetric EFT. In this sense, Conjecture \ref{conj:BPS} can be understood as the equality between \eqref{L} and \eqref{Lsm} for the case of D-branes that wrap torsion cycles and that at the same time are mutually BPS, that is, they preserve some common supercharges in the lower-dimensional EFT.      Notice that equating \eqref{L} to \eqref{Lsm} implies a cancellation in the contribution of very massive modes to the torsion linking number. Physically this suggests that a protection mechanism against threshold corrections must be in place, which is indeed a characteristic feature of certain supersymmetric settings.

To make the proposal more precise, a number of comments are in order. First, some of the compactification manifolds that we will consider correspond to supersymmetric $D$-dimensional EFTs without vacua in the interior of their field space. Instead, they describe supersymmetric solutions that probe a family of metrics of $X_n$. For this reason, we require that the torsion representatives that are BPS/calibrated must be so in a region of the EFT field space, as opposed to in a single point. In particular, they must remain calibrated upon local deformations of the metric that either are moduli or involve energies below the compactification scale. Notice that, in general, calibrated $p$-cycles in fixed homology classes can cross walls of marginal or threshold stability when one deforms the metric of the compactification manifold, so this condition is a significant restriction in the definition of calibrated submanifolds, that we will dub strict calibration condition. In the string theory literature, examples of BPS objects with this property are the EFT strings and membranes defined in \cite{Lanza:2020qmt,Lanza:2021udy,Lanza:2022zyg}, so in this sense some the objects of study in this work can be thought of as their torsion analogues. 

Second, notice that Conjecture \ref{conj:BPS} implies that the smeared linking number should not vary upon infinitesimal deformations of the embedding of the torsion representatives that respect the calibration condition, which we will dub as BPS deformations. In the following sections we argue that this is indeed the case, by relating the coefficients $c_i$, $e_i$ with the volume of the respective $p$-cycles. Now, since we are interested in metrics with non-closed calibration $p$-forms, one may in principle encounter BPS deformations that vary the $p$-cycle volume. We have not found instances of this possibility for our more restrictive definition of calibrated cycle. However, in case that it occurred to apply Conjecture \ref{conj:BPS} one should consider the calibrated representative of $\Pi_p$ that locally minimises its volume for a fixed metric in $X_n$. 

Finally, using the bilinearity of the linking number one may extend the conjecture to torsion $p$-cycles that are not calibrated by themselves, but that are linear combinations of calibrated cycles. For instance, let us consider a manifold $X_n$ with a $G$-structure metric and a pair of $p$-cycles $\Pi_p$ and $\Pi_p'$ calibrated by the same calibration, both in the strict sense, that correspond to the same class on $H_p(X_n, \IR)$, but such that $[\Pi_p^{\rm tor}] = [\Pi_p'] - [\Pi_p]$ is a non-trivial element of ${\rm Tor} H_p(X_n, \IZ)$. Then one may smear out both delta-forms separately, and define a smeared description of the torsion two-cycle as
\be
\delta^{\rm sm} (\Pi_p') - \delta^{\rm sm} (\Pi_p) \, .
\label{deldif}
\ee
By construction, this is an exact smooth $(n-p)$-form, from where one can extract the coefficients $c_i$ as in \eqref{deltasm}. A different realisation of $\Pi_p^{\rm tor}$ in terms of BPS cycles, like for instance a representative $\Pi_p^{\rm tor}$ that is BPS by itself, may give rise to different coefficients $c_i$. However, the extension of the conjecture would imply that all these choices give rise to the same smeared linking number with a given BPS torsion $(n-p-1)$-cycle. Notice that with this extension one may not only compute torsion linking numbers via smeared data in $G$-structure manifolds with non-closed calibrations, but also in manifolds with metrics of special holonomy. 

To sum up, our proposal means that for manifolds endowed with certain metrics, one can compute some torsion invariants in terms of smeared/massive EFT data. One only needs {\it i)} the eigenforms of the Laplacian that correspond to their lowest eigenvalues and {\it ii)} the projection of torsion, strict-calibrated cycles into them.

%%%%%%%%%%%%%%%%%%%
%%%%%%%%%%%%%%%%%%%

\section{Localised sources and dimensional reduction}
\label{s:dimred}

The aim of this section is to motivate the content of Conjecture \ref{conj:BPS} from a physics viewpoint, by considering the effect of localised sources in compactifications of string theory. If these sources wrap torsion cycles in the compact dimensions and couple to the massive fields present in the lower-dimensional EFT, then by consistency of the low-energy description there should be terms in the EFT Lagrangian that know about their torsion linking number. The reason is that, in this case, the EFT contains localised objects charged under a discrete gauge symmetry (the torsion cohomology group) with a non-trivial backreaction at EFT wavelengths. The corresponding EFT Lagrangian has the form proposed in \cite{Camara:2011jg} (see also \cite{Grimm:2011tb,Berasaluce-Gonzalez:2012abm,Berasaluce-Gonzalez:2012awn,Mayrhofer:2014laa,Grimm:2015ona,Braun:2017oak}) to describe torsion in (co)homology from the viewpoint of dimensional reduction. However, this does not guarantee that one can compute the torsion linking number from smeared data. For this, one in addition needs that such localised sources appear as BPS objects of the EFT. 

\subsection{Localised sources in ten and four dimensions}

For concreteness, let us consider a static D4-brane in 10d, with worldvolume $\Sigma_5 = \IR \times \Sigma_4 \subset \IR^{1,9}$. Its backreaction sources a RR field strength $F_{4} = dC_{3}$, such that $dF_{4} = \delta_5(\Sigma_5)$  corresponds to the bump delta 5-form with support on $\Sigma_5$. On a 4-sphere $S^4$ surrounding this source, the pullback of $F_4$ is of the form $2\pi \Phi_{S^4}$, where $\Phi_{S^4}$ is such that $\int_{S^{4}} \Phi_{S^4} = 1$. Analogously to the Wu-Yang description of a 4d monopole, we need to describe the potential $C_{3}|_{S^4}$ as a connection, more precisely as the connection of a 2-gerbe on $S^4$, see e.g. \cite{Hitchin:1999fh}. 

Let us  recall how a probe D2-brane feels this background. In particular, let us consider the case where its worldvolume $\Sigma_{3}$ sweeps a 3-sphere $S^{3}$ at the equator of $S^{4}$. In analogy with the Wu-Yang monopole, the non-trivial pull-back of $F_{4}$ on $S^{4}$ has the effect that $C_{3}|_{S^{3}}$ is non-trivial in the cohomology of $S^3$. However, one can still globally write it as $d\lambda_{2}$, where $\lambda_{2}$ is not a globally well-defined smooth two-form, but nevertheless $e^{i\int_{\Pi_{2}} \lambda_{2}}$ is well-defined for any two-cycle $\Pi_{2}$ inside  $S^{3}$. This property amounts to saying that the wavefunction of the probe D2-brane is well-defined in the backreacted background of its magnetic dual. 

We now consider the particular 10d background $\IR^{1,3} \times X_6$, with $X_6$ a compact manifold with a given metric.  If the D4-brane wraps a 4-cycle $\Pi_4 \subset X_6$, then it will look like a point-like source in the 4d EFT, very much like a monopole. Its backreaction may be described by a 2-gerbe in the microscopic 10d picture, but its effective description at wavelengths larger than $1/m_{\rm KK}$  should correspond to a bundle similar to that of the Wu-Yang monopole. This is indeed the case whenever $[\Pi_4]$ is a non-trivial class in $H_4 (X_6,\IR)$. A probe D2-brane wrapping a two-cycle $\Pi_{2} \subset X_6$ with non-trivial transverse intersection $ \Pi_2 \cdot \Pi_{4} = Q$ looks, from the 4d viewpoint, like a (test) particle circling around the monopole-like source. The pull-back of $F_{4}$ on $S^2  \times \Pi_{2}$ with the two-sphere surrounding the source reads:
\be
2\pi \left( \Phi_{S^2} + d\chi_1\right) \wedge \delta_2(\Pi_4)|_{\Pi_{2}} = 2\pi Q \left( \Phi_{S^2} + d\chi_1\right) \wedge \left(\Phi_{\Pi_{2}} + d\tilde{\chi}_{1}\right)  ,
\label{pullF84d}
\ee
where the $\Phi$'s are volume forms normalised to unity, $\chi_1$ and $\tilde{\chi}_1$ are globally well-defined one-forms on $S^2$ and $\Pi_4$,  respectively, and $
\delta_2(\Pi_4)$ is the bump delta two-form of $\Pi_4$ in $X_6$. The result gives an integral of $2\pi Q$, that corresponds to the product of electric and magnetic charges. We can now restrict our attention to $\gamma \times \Pi_{2}$, where $\gamma$ is at the equator of $S^2$. The difference of connections $C_{3}$ on two patches overlapping over  this submanifold can be written as  
\be
C_{3} = d\lambda_{2}, \qquad \text{with} \quad \lambda_{2} = \lambda\, Q \left( \Phi_{\Pi_{2}} + d\tilde{\chi}_{1}\right) ,
\label{C3point}
\ee
and $\lambda$ a function $\lambda : \gamma \to S^1$ with a single winding. When we describe this system at energies well below the compactification scale, we simplify the internal profile for $\lambda_{2}$. In the effective description, one replaces $ \delta_2(\Pi_4)$ by a harmonic two-form $\omega_2^{\Pi_4}$ in the Poincar\'e dual class to $[\Pi_4]$, which is the lowest lying mode (the harmonic piece) of the Kaluza--Klein (KK) decomposition of $\delta_2(\Pi_4)$. Therefore, we write $ \lambda_{2} = \lambda Q \omega_2^{\Pi_4}$, as a more detailed profile would involve gauge transformations for massive $U(1)$'s that are beyond our 4d EFT description. 

If we now assume that $[\Pi_4] \in {\rm Tor}H_4(X_6, \IZ)$, then $\omega_2^{\Pi_4}$ vanishes, and the presence of a D4-brane wrapped on $\Pi_4$ remains undetected by any D2-brane wrapping a two-cycle $\Pi_2 \subset X_6$. Instead, as discussed in \cite{Camara:2011jg}, one needs to consider a D4-brane wrapping  a three-cycle $[\Pi_3] \in {\rm Tor}H_3(X_6, \IZ)$, which is perceived by the low-energy EFT as a 4d string. Let us for simplicity place this 4d string in $\IR^{1,1} \subset \IR^{1,3}$, and take $z = r e^{2\pi i \theta}$ to be the complex coordinate transverse to its worldsheet. This time one can provide a global description of the RR potential sourced by the D4-brane
\be
C_{3} = 2\pi d \left( \theta \rho_{2} \right) ,
\label{Cstringb}
\ee
where $\rho_2$ is a two-form on $X_6$ such that $d\rho_2 = \delta_3(\Pi_3)$ is the bump delta three-form of $\Pi_3$. This background is detected by a D2-brane with worldvolume $\g \times\Pi_2$, where $[\Pi_2] \in  {\rm Tor} H_2(X_6, \IZ)$ is a torsion class with linking number $L \in \mathbb{Q}$ with respect to $[\Pi_3]$, and $\gamma$ is a 4d worldline surrounding once the string location $\{z=0\}$. If we pull back \eqref{Cstringb} into the D2-brane worldvolume we obtain
\be
C_{3}|_{\g \times \Pi_{2}} = 2\pi L \left( d\phi + df(\phi) \right) \wedge \left(\Phi_{\Pi_{2}} + d\tilde{\chi}_{1}\right)  ,
\label{pullC74d}
\ee
where $\phi \in \IR/ \IZ$ parametrises $\gamma$ and $f(\phi)$ is a periodic function in it, and we have used that $\int_{\Pi_{2}} \rho_{2} = L$. Therefore, we obtain that  $C_{3} = d\lambda_2$, with $\lambda_{2}$ of the form \eqref{C3point}, except for the replacement $Q \to L$. The fact that $L$ is not an integer number implies that the D2-brane picks a non-trivial phase $e^{2\pi i L}$ when circling around $\gamma$, which is a trait of 4d Aharanov-Bohm (AB) strings and signals the presence of a discrete gauge symmetry \cite{Banks:2010zn}. 

Let us describe the discrete gauge symmetry in terms of the gauge transformations involved in the backreacted D4-brane background. At the microscopic 10d level, these are of the form
\be
d \left( \lambda \rho_{2}\right) = d\lambda \wedge  \rho_{2} + \lambda \, \delta_3 (\Pi_{3}) ,
\label{truel}
\ee
with $\lambda$ well-defined on loops on $\IR^{1,3}$ but not on $\IR^{1,3}$ itself. It now remains to see what is the long-wavelength 4d EFT description of this transformation. As already discussed $\delta_3 (\Pi_{3})$ has no harmonic component, and the same can be assumed for $\rho_2$.\footnote{A priori nothing forbids $\rho_2$ to have a  harmonic piece, which would even be required if we impose that $\int_{\Pi_2} \rho_2 \in \IZ$ for any two-cycle $\Pi_2$ \cite{Marchesano:2014iea}. Following \cite{Marchesano:2014bia}, this piece would imply a non-trivial kinetic mixing between massive and massless $U(1)$'s of the compactification, which could then be removed by an  appropriate change of basis. To simplify the discussion, here we assume the absence of such a harmonic piece.} The question is then if $\delta_3 (\Pi_{3})$ has a non-trivial projection into the massive field content of the 4d EFT spectrum, or in other words if it has a non-trivial 4d smearing. If it does, the D4-brane backreaction should be seen by the 4d EFT, in the sense that it sources some of it fields, that pick a non-trivial profile involving wavelengths above $1/m_{\rm KK}$. So in the following we will assume that $\delta_3 (\Pi_{3})$ has a non-trivial 4d smearing, which is also necessary for Conjecture \ref{conj:BPS} to provide a non-trivial statement.

For simplicity let us assume that $X_6$ is such that ${\rm Tor}H_3(X_6,\IZ) =\IZ_N$. By the Universal Coefficient Theorem \cite{Bott1982DifferentialFI} and Poincar\'e duality this implies that ${\rm Tor}H_2(X_6,\IZ) =\IZ_N$, and also that $LN \in \IZ$. Let us in addition assume that there is a single exact eigen-three-form $b_3$ of the Laplacian with unit norm and a non-vanishing eigenvalue below the compactification scale. That is, we have a unique solution of the form $dd^{\dag} b_3 = \lambda_{\rm st}^2 b_3$, with $m_{\rm st} = \ell_s^{-1}\lambda_{\rm st} \ll m_{\rm KK}$. Then to obtain our 4d EFT via dimensional reduction we must consider the following set of $p$-forms
\be
b_3 , \qquad *b_3, \qquad  \lambda_{\rm st}^{-1} d* b_3,  \qquad \lambda_{\rm st}^{-1} d^\dag b_3 ,
\label{massivepforms}
\ee
all of them with unit norm and the same eigenvalue, because they are associated with the same mass scale. The standard dimensional reduction procedure consists of expanding the 10d $p$-form potentials in a basis of harmonic forms plus the above, non-harmonic set. For instance, reducing the type IIA three-form $C_3$ to 4d with respect to the above non-harmonic sector gives 
\be
C_3 = 2\pi  \ell_s^{3}  \left(A_1 \wedge \om_2 +  C_0 \, \b_3 \right) ,
\label{C3sing}
\ee
where $A_1$ and $C_0$ describe a 1-form and a 0-form in 4d, respectively, and we have defined
\be
\b_3 = f\, b_3 , \qquad \om_2 =  \frac{1}{g \lambda_{\rm st}} d^\dag b_3 , \qquad \text{with} \quad f, g \in \IR .
\label{alom}
\ee
so $d\om_2 = N_{\rm eff} \b_3$ with $N_{\rm eff} = \frac{\lambda_{\rm st}}{fg}$. The reduction to 4d of the 10d kinetic term $\int F_4^2$ gives
\be
(2\pi  \hat{f})^2 \left( dC_0 - N_{\rm eff} A_1 \right)^2 + \frac{4\pi^2}{g^{2}} (dA_1)^2 , \qquad \hat{f} \coloneqq  f e^{\phi_4} M_{\rm P} = \frac{f e^{\phi}}{(4\pi{\rm Vol}_{X_6})^{1/2}} M_{\rm P}  ,
\label{stuck}
\ee
namely a St\"uckelberg-like Lagrangian, where $e^{\phi_4}$ is the 4d and $e^{\phi}$ the 10d dilaton, Vol$_{X_6}$ is the volume of $X_6$ in string units, and $M_{\rm P}$ the 4d Planck mass. This is precisely the dimensional reduction scheme proposed in \cite{Camara:2011jg} to describe discrete gauge symmetries from torsion in cohomology, if one imposes the constraint $N_{\rm eff} = N$ and treats $C_0$ as an axion-like particle of unit periodicity $C_0 \sim C_0 +1$. In this case, the discrete gauge symmetry is generated by the shift
\be
2\pi C_0 \to 2\pi C_0 +\lambda , \qquad 2\pi A_1 \to  2\pi A_1 +  \frac{d\lambda}{N} ,
\label{disgauge}
\ee
with $\lambda \in 2\pi \IZ$. A particle with charge $NL$ under $A_1$ will pick up a phase $e^{2\pi i L}$ upon \eqref{disgauge}, for instance when circling a string of unit charge. This is how the 4d EFT reflects the linking number between torsion cycles on $X_6$, and in particular that ${\rm Tor} H_3(X_6,\IZ) \simeq $ ${\rm Tor} H_2(X_6,\IZ) =\IZ_N$. So while at this point we have not determined the parameters $f$ and $g$, consistency of the 4d EFT requires that they are constrained by $fg = \frac{\lambda_{\rm st}}{N}$. Therefore we have the relation
\be
\frac{1}{N} = \frac{fg}{\lambda_{\rm st}} = \frac{\hat{f}g}{m_{\rm st}} .
\label{Lphys}
\ee
Note that the expression in the middle resembles a smeared linking number, as defined in \eqref{L}, while the rhs corresponds to how the  EFT massive sector encodes this quantity.

The 4d effective Lagrangian \eqref{stuck} should be sufficient to give a long-wavelenght description (more precisely in the range $(m_{\rm KK}^{-1}, m_{\rm st}^{-1}))$ of the backreaction of D4-branes wrapping torsion three-cycles of $X_6$. Recalling our 10d analysis, one may try to provide such a 4d description by directly smearing the 10d solution, that is by projecting the background \eqref{Cstringb} into the massive sector \eqref{massivepforms}. However, if one does so the gauge transformation \eqref{truel} translates into
\be
2\pi C_0 \to 2\pi C_0 + \frac{c}{f} \lambda , \qquad 2\pi A_1 \to  2\pi A_1 + \frac{c}{f}  \frac{d\lambda}{N} ,
\label{disgauge2}
\ee
where $\delta_3^{\rm sm}(\Pi_3) = c b_3$, and we have imposed that $N_{\rm eff} = N$. So only when $c=f$ we recover the expected gauge transformation \eqref{disgauge}. While this may seem surprising, it does not necessarily indicate any inconsistency of the 4d EFT. Instead, one may interpret it as the fact that smearing a 10d background is a classical procedure that may be subject to corrections, like quantum corrections associated to the fields above the compactification scale that one has truncated.   So in principle, it could be that these or other corrections modify the backreacted 4d background in such a way that the quotient $c/f$ disappears from \eqref{disgauge2}, and one recovers the gauge transformation \eqref{disgauge} consistent with \eqref{stuck}. If this was the case, \eqref{Lphys} should be interpreted as a smeared linking number after corrections have been taken into account. 

This proposal to solve the apparent inconsistency in \eqref{disgauge2} has the downside that it does not give a clear geometric prescription to compute the parameters $f$ and $g$ which, together with $\lambda_{\rm st}$, are the 4d EFT data that allow us to compute $N$. However, it gives us the guideline that one should try to consider D4-branes whose smeared backreaction does not suffer important corrections upon dimensional reduction. From a physics viewpoint, the best candidates to display this feature are D4-branes that preserve some supersymmetry of the background, namely BPS objects of the EFT, as the results of \cite{Blaback:2010sj} also suggest. In the next subsection we will argue why this is the right answer. 

Finally, it is instructive to perform the dimensional reduction of the RR potential $C_5$, dual to $C_3$ in 10d. An expansion in the relevant non-harmonic $p$-forms \eqref{massivepforms} gives
\be
 C_{5} = 2\pi \ell_s^{5}\left( V_1 \wedge  \tilde{\omega}_4 +  B_2 \wedge \a_3 \right) , 
\label{C5sing}
\ee
where $V_1$ and $B_2$ are a 4d 1-form and 2-form in 4d, and we have defined
\be
 \tilde{\omega}_4  = \frac{g}{\lambda_{\rm st}} d * b_3 , \qquad \a_3 =  f^{-1} * b_3  .
 \label{tombe}
\ee
such that $\int_{X_6} \omega_2 \wedge \tilde{\omega}_4 = \int_{X_6} \a_3 \wedge \b_3 = 1$, as in \cite{Gurrieri:2002wz,DAuria:2004kwe,Grana:2005ny,Kashani-Poor:2006ofe}.  This is required for the fields $(V_1, B_2)$ to be quantised 4d duals to $(A_1, C_0)$. It also implies that $d\a_3 = N_{\rm eff} \tilde{\om}_4 = N \tilde{\om}_4$, so upon dimensional reduction one obtains 
\be
 (2\pi g)^2 \left( dV_1 + N B_2  \right)^2 + \frac{4\pi^2}{\hat{f}^{2}}  (dB_2)^2 .
\label{stuckdual}
\ee
This 4d effective Lagrangian should describe the backreaction of D2-branes wrapping torsion two-cycles $\Pi_2$ of $X_6$, in the long-wavelength approximation. In these dual variables the discrete gauge symmetry reads
\be
2\pi V_1 \to 2\pi V_1 - \lambda_1 , \qquad 2\pi B_2 \to  2\pi B_2 +  \frac{d\lambda_1}{N} .
\label{disgaugedual}
\ee
A D2-brane wrapping a torsion two-cycle $\Pi_2$ such that $\delta_4^{\rm sm}(\Pi_2) = \frac{e}{\lambda_{\rm st}} d* b_3$ will not generate this shift via its backreaction, unless $e = g$. Again, one could interpret this mismatch as the result of non-trivial quantum corrections, and argue that the equality should hold for D2-branes wrapping calibrated two-cycles, as we proceed to argue.

\subsection{The supersymmetric case}
\label{ss:BPS}

One may summarise the reasoning of the previous subsection as follows. A D4-brane wrapping a torsion three-cycle $\Pi_3$ of $X_6$ becomes, upon compactification to 4d, an Aharanov-Bohm 4d string that realises ${\rm Tor} H_3(X_6, \IZ) = \IZ_N$ as a discrete gauge symmetry. This object will be perceived by the 4d EFT if it couples to some massive $p$-form modes below the compactification scale $m_{\rm KK}$. In that case the backreaction has size $m_{\rm st} = \lambda_{\rm st}/\ell_s \ll m_{\rm KK}$, where $\lambda_{\rm st}$ is the eigenvalue of such massive eigenmodes, and there must be a term in the 4d EFT that describes such a backreacted solution at long wavelengths. This 4d Lagrangian term is \eqref{stuck}, with $N_{\rm eff} = \lambda_{\rm st}/fg = N$ encoding the topological information of the torsion homology group. Knowledge of the massive spectrum and of the parameters $f, g \in \IR$ thus allows us to compute torsion cohomology groups, and to represent them via smooth $p$-forms \eqref{alom} and \eqref{tombe} that are, from the 4d viewpoint, analogous to the harmonic representatives of de Rham cohomology groups. The parameters $f$ and $g$ are not determined from the 4d smearing of the backreaction of a D4-brane wrapping an arbitrary torsion three-cycle, since in general there can be significant corrections to the smeared background. Notice that these parameters are intrinsic of the 4d EFT, and so they only depend on the topology and metric of $X_6$. 

There is however a particular class of 4d strings for which quantum corrections should be under control, namely BPS fundamental strings of the EFT. We are particularly interested in D4-branes that correspond to the 4d EFT strings of \cite{Lanza:2020qmt,Lanza:2021udy,Lanza:2022zyg}, except that they source 4d axions with a mass $m_{\rm st}$ induced by a St\"uckelberg coupling. As stressed in \cite{Lanza:2020qmt}, near the string core and at wavelenghts below $m_{\rm st}^{-1}$ one should be able to describe the 4d backreaction of these objects with a solution similar to that of standard EFT strings, implying that their tension is determined by the kinetic terms of the 4d EFT Lagrangian, and in particular by parameters like $f$. 

Geometrically, the BPS condition means that the torsion three-cycle $\Pi_3$ is calibrated by a complex three-form $\Omega$. This is not possible when $X_6$ is a Calabi--Yau, but it occurs in SU(3)-structure manifolds with a metric specified by $(J, \Omega)$ and a non-vanishing intrinsic torsion,\footnote{The two meanings of the word torsion should not be confused. By intrinsic torsion we mean the five torsion classes which enter in the description of manifolds with SU(3)-structure metrics, and which show up in the derivatives of the globally well-defined forms $\Omega$ and $J$ \cite{Gray1980TheSC,chiossi2002intrinsic,Hull:1986iu,Strominger:1986uh,LopesCardoso:2002vpf,Gauntlett:2003cy,Grana:2005jc,Koerber:2010bx}. In any other instance, the word torsion refers to torsion classes in (co)homolgy groups of $X_6$, and to their representatives.} which we will assume in the following. Notice that the calibration condition selects a specific representative within the torsion class $[\Pi_3] \in {\rm Tor} H_3(X_6, \IZ)$, that directly depends on the metric of $X_6$. Therefore, it is reasonable to assume that $f$, which also depends on the metric of $X_6$, can be computed from $\delta_3(\Pi_3)$ with $\Pi_3$ calibrated. More precisely, we will argue that $f$ can be computed from the smeared delta-form $\delta_3^{\rm sm}(\Pi_3)$.

To see this, let us assume that upon compactification of type IIA string theory on $X_6$ we recover a 4d EFT with $\CN= 2$ supersymmetry. According to \cite{Grana:2005ny,Kashani-Poor:2006ofe} this is guaranteed if $X_6$ is an SU(3)-structure manifold with calibrations $(J,\Omega)$. Following their approach, we may expand $J$ and $\Omega$ in the set of harmonic two- and three-forms, respectively, plus the non-harmonic set \eqref{massivepforms}. Let us first consider $\Omega$ and assume an expansion of the form $\Omega = \Omega^{\rm harm} + i a \,  \a_3 +  b \, \b_3$, where $\Omega^{\rm harm}$ is a sum of harmonic three-forms and $a$, $b$ are real functions of the 4d fields. If we impose the condition $* \Omega = - i \Omega$ we find that the more precise form 
\be
\Omega = \Omega^{\rm harm} + {\rm Vol}_{X_6}^{1/2} \left(if \a_3 +  f^{-1} \b_3 \right) ,
\ee
where we have taken into account that $\int_{X_6} i \bar{\Omega} \wedge \Omega = 8 {\rm Vol}_{\rm X_6}$, and that $\a_3, \b_3$ are orthogonal to any harmonic form. Notice that here $f$ is not a fixed number, but depends on the choice of SU(3)-metric or, from the 4d viewpoint, on the vevs of the 4d scalar fields.

Next, we use that for a 4d BPS  string its tension is proportional to the  kinetic term of the axion to which it couples magnetically. In the  case at hand, the orthogonality of the massive modes implies that the string charge-to-mass ratio equals one, and so for a string inducing a single winding of $C_0$ around its core the tension is determined by the axion decay constant in \eqref{stuck} as ${\sqrt{\pi}} \hat{f}M_{\rm P}$. This quantity should correspond to the 4d string tension obtained from a D4-brane wrapped on a three-cycle $\Pi_3$ calibrated by $\Omega$, see \cite[section 6.4]{Lanza:2021udy}. We thus find
\be
\frac{\hat{f}}{M_{\rm P}}= \frac{e^{\phi}}{{\sqrt{4\pi}} {\rm Vol}_{X_6}} \left| \int_{\Pi_3}  \Om \right| \implies  f = {\rm Vol}_{X_6}^{-1/2} \left| \int_{X_6}  \Om \wedge \delta_3(\Pi_3) \right| .
\label{fst}
\ee
Notice that in the second equation we can replace $\delta_3(\Pi_3) \to \delta_3^{\rm sm}(\Pi_3)$. Using that $\Pi_3$ is a torsion three-cycle and therefore $\delta_3(\Pi_3)$ is an exact three-form we finally obtain
\be
\delta_3^{\rm sm}(\Pi_3) = \b_3 .
\ee
That is, $f$ can be found from smearing the bump delta-form of a calibrated torsion three-cycle. 

Similarly, one may consider a D2-brane wrapping a BPS representative of $[\Pi_2] \in {\rm Tor} H_2 (X_6, \IZ)$, or in other words $\Pi_2$ is calibrated by $J$. A BPS particle of unit charge with respect to $A_1$ will have a mass $g M_{\rm P}$, so putting both statements together results in the equality
\be
g = {\rm Vol}_{X_6}^{-1/2} \left| \int_{X_6}  J \wedge \delta_4(\Pi_2) \right| .
\label{gpt}
\ee
Again, expanding $e^{iJ}$ in harmonic and non-harmonic forms and using the Hodge duality relations translates into the equality 
\be
\delta_4^{\rm sm}(\Pi_2) = \tilde{\om}_4 .
\ee
Equivalently, $g$ results from smearing the bump delta-form of a calibrated torsion two-cycle.

Notice that in this construction the torsion cycles $\Pi_3$ and $\Pi_2$ that lead to $f$ and $g$ have a minimal 4d charge and tension. Therefore we expect them to generate ${\rm Tor} H_3(X_6, \IZ)$ and ${\rm Tor} H_2(X_6, \IZ)$, respectively, and to have a linking number $1/N \mod 1$. When plugging the values of $f$ and $g$ into the smeared linking number one indeed finds that $L^{\rm sm} (\Pi_2, \Pi_3) =1 /N$, in agreement with Conjecture \ref{conj:BPS}. If instead $\Pi_2$ corresponds to a particle of charge $LN$, then repeating the same reasoning its smeared delta-form will have to be multiplied by $LN$, and we will recover a smeared linking number of $L$, again supporting the conjecture. 

An interesting point is that, when dealing with mutually BPS objects, one should be able to add up their tensions to compute the energy of the total system. Geometrically, this amounts to say that even if the topological charge of a calibrated torsion $p$-cycle or a sum of them lives in $\IZ_N$, its central charge lives in a lattice. This does not imply any contradiction with the $\IZ_N$ discrete system of the 4d EFT, provided that the process that reduces the number of BPS objects by $N$ has a non-vanishing energy which compensates for the loss of $N$ $p$-cycles. To illustrate this, let us consider the torsion three-cycle class $[\Pi_3]$ generating ${\rm Tor} H_3(X_6, \IZ)$ in a SU(3)-structure manifold. A set of $N$ D4-branes wrapping calibrated representatives $\Pi_{3,i}$ of this class looks like $N$ BPS strings in 4d. These can end on a 4d monopole, made up of a D4-brane wrapping a four-chain $\Sigma_4$ whose boundary is given by $\p \Sigma_4= \sum_i \Pi_{3,i}$ \cite{Camara:2011jg}. On the one hand, using Stokes' theorem one can relate the sum of string tensions with the integral of $d\Omega$ over $\Sigma_4$. On the other hand, the mass of the 4d monopole is proportional to the volume of $\Sigma_4$ which, if the monopole is BPS, is given by the integral of $\pm \oh J \wedge J$ over $\Sigma_4$. Therefore we find that the marginal stability of $N$ BPS AB strings implies
 \be
\ell \left| \int_{X_6} d\Omega \wedge \delta_2(\Sigma_4)\right| + \left| \oh \int_{X_6} J \wedge J  \wedge \delta_2(\Sigma_4) \right| = {\rm const.} 
\label{BPSmono}
\ee
where $\ell$ is the length of the AB string in $\ell_s$ units.  Notice that this relation can only make sense if the monopole mass depends on $\ell$, which should then be a feature of  backgrounds with BPS AB strings and particles. We postpone a more precise explanation of this statement to the next section, where both quantities in \eqref{BPSmono} will be evaluated in a simple setup based on half-flat manifolds. For the time being, it is worth pointing out that the above reasoning leads to an interpretation of the non-closed two-form $\om_2$ in \eqref{alom}. Indeed, notice that in \eqref{BPSmono} we can replace $\delta_2(\Sigma_4) \to \delta_2^{\rm sm}(\Sigma_4)$ and that because the action of smearing commutes with the exterior derivative, $d \delta_2^{\rm sm}(\Sigma_4) = N \delta_3^{\rm sm}(\Pi_3) = N \b_3$. It is thus natural to guess that $\delta_2^{\rm sm}(\Sigma_4) = \om_2$ when $\Sigma_4$ is a calibrated four-chain, something that can be verified by noting that a 4d BPS monopole of unit charge has mass $g^{-1} M_{\rm P}$, and running a reasoning analogous to the previous ones. Similarly, one can deduce that $\delta^{\rm sm}_3(\Sigma_3) = \a_3$, where $\Sigma_3$ is a calibrated three-chain ending on $N$ calibrated torsion two-cycles. Therefore, one concludes that the set of harmonic plus non-harmonic forms in which one expands $J$, $\Omega$ and the RR potentials to obtain the 4d fields can be interpreted as smeared delta-forms of a basis of calibrated chains and cycles. Notice that this fits well with the notion that the set of forms $\{ \om_2, \a_3, \b_3, \tilde{\om}_4\}$ reflect quantisation features of the 4d EFT, like axions of unit periodicity and $U(1)$ gauge symmetries. This quantisation also implies that these $p$-forms generate a lattice just like quantised harmonic $p$-forms do, which seems to be in contradiction with the fact that these D-brane charges are torsion. However, as mentioned above when dealing with mutually BPS objects the mass/tensions are additive, which explains the lattice structure. Finally, while here we have considered a very simple case, it is reasonable to expect that this description of the reduction basis of $p$-forms extends to the general framework of SU(3)-structure manifold dimensional reduction analysed in \cite{Grana:2005ny,Kashani-Poor:2006ofe}, as we will further discuss in section \ref{s:general}. 

It is also instructive to consider what happens when we slightly depart from a BPS embedding. In particular, let us take a D4-brane wrapping a calibrated torsion three-cycle $\Pi_3$ and perform a small deformation of its embedding, such that the torsion linking number with a calibrated torsion two-cycle $\Pi_2$  does not change
\be
L( \Pi_{2},\Pi_3) = \sum_i \frac{c_ie_i}{\lambda_i}  \, .
\label{Labs}
\ee
Notice that this quantity is not defined mod 1. Geometrically, this means that upon the deformation $\Pi_3$ does not cross $\Pi_2$. A simple deformation of this sort that changes the smeared linking number takes the form 
\be
c_{\rm KK} \to c_{\rm KK} - \eps , \qquad c_{\rm st} \to c_{\rm st} + \frac{\lambda_{\rm st}}{\lambda_{\rm KK}} \eps ,
\label{shiftKK}
\ee
where $c_{\rm KK}$ represents the coefficient of a mode above the compactification scale and $c_{\rm st}$ one below. We thus find that the naive gauge transformation \eqref{disgauge2} changes with a suppression factor of  $m_{\rm st}/m_{\rm KK}$ with respect to the BPS case. This is indeed the kind of suppression that one would expect from integrating out massive operators at the Kaluza-Klein scale, which supports the interpretation that the expected discrete gauge transformation \eqref{disgauge} could be restored for the non-BPS case, once that quantum corrections are taken into account. It would however be important to perform a more direct test of this proposal.

\subsection{Generalisations}

In our discussion so far we have focused in a type IIA setup, in which D4- and D2-branes look respectively like strings and particles in the 4d EFT. However, it is clear that the same reasoning can be applied to any other kind of string compactifications, as long as the 4d picture is similar. For instance, in type IIB compactified in a SU(3)-structure manifold, Aharanov-Bohm strings and particles would be realised by D3-branes wrapping torsion two- and three-cycles. There are other extended objects that can give rise to 4d AB strings and particles \cite{Camara:2011jg}, but in many instances they do not wrap calibrated cycles, and so the BPS property, which is an important ingredient of our logic, is missing.

Nevertheless, one may extend our reasoning in yet another direction, since there are other BPS objects in a 4d EFT that encode torsion in cohomology. Indeed, a key property of AB strings with $\IZ_N$ charge is that $N$ of them can end on a monopole, while $N$ AB particles can end on a 4d instanton \cite{Banks:2010zn}. As a general rule, $\IZ_N$ charges are detected in the 4d theory by $p$-branes ending on $(p-1)$-branes with $p = 0,1,2,3$, and in certain instances these $\IZ_N$ charges reflect torsion cohomology groups of the compactification manifold \cite{Berasaluce-Gonzalez:2012awn}. In our previous discussion we have focused on the cases $p=0$ and $p=1$, which are typically represented in 4d EFT language by the Lagrangians \eqref{stuck} and \eqref{stuckdual}, respectively, and are dual to each other. The case $p=2$ corresponds to 4d membranes ending on strings, and it is related to the following EFT Lagrangian piece \cite{Marchesano:2014mla}
\be
\left( dB_2 - N C_3\right)^2 ,
\label{p=2}
\ee
where $C_3$ is a three-form that couples to the membrane and $B_2$ is a two-form coupling to the string. In our previous type IIA setup, these objects would arise from D4-branes wrapping torsion two-cycles and the three-chain connecting them, respectively, and signal the presence of a non-trivial superpotential. The case $p=3$ describes 4d space-time filling branes ending on membranes, and the corresponding Lagrangian piece reads \cite{Lanza:2019xxg}
\be
\left( dD_3 - N A_4\right)^2 ,
\label{p=3}
\ee
where $A_4$ coupling to the space-time filling branes and $D_3$ to the membranes. In our type IIA setup these 4d objects arise from D6-branes wrapped on torsion three-cycles and on a four-chain linking them, respectively.\footnote{In most of the literature, these Lagrangians are shown to arise from compactifications with NS $H$-fluxes. In this case, $N$ represents an $H$-flux quantum and the feature of $p$-branes ending of $(p-1)$-branes has a microscopic description in terms of $d_H$ cohomology and its dual homology \cite{Evslin:2007ti}. By looking at concrete constructions, it is easy to convince oneself that such a setup is connected by mirror symmetry to the one that we are considering \cite{Tomasiello:2005bp,Marchesano:2006ns}.} 

The general philosophy of the previous subsection also applies to these St\"uckelberg-like couplings. That is, the 4d $p$-forms that appear in \eqref{p=2} and \eqref{p=3} should arise from expanding the 10d RR potentials on smeared delta-forms of calibrated torsion cycles of $X_6$. The resulting coefficients that multiply both expressions are the analogues of $f$ and $g$ in \eqref{stuck} and \eqref{stuckdual}, and so together with the relevant Laplace eigenvalue they determine $N$. Notice that the analogy is not straightforward, because  \eqref{p=2} and \eqref{p=3}  are not dual Lagrangians, which reflects the fact that D4-branes and D6-branes do not couple to dual 10d RR potentials. However one may consider a gauge instanton on the space-time filling D6-branes, which amounts to a D2-brane wrapping the same torsion three-cycle and coupling to a massive 4d axion $C_0'$ that arises from reducing $C_3$ on a coexact three-form like $\a_3$ in \eqref{tombe}. This EFT object is sensitive to the backreaction of D4-branes wrapping torsion two-cycles and the three-chain connecting them, and an analogy with the gauge transformations involving AB strings and particles can be drawn. The precise statement is that there exists a gauged $(-1)$-form symmetry that describes the discrete gauge symmetries of the EFT superpotential \cite{Hebecker:2017wsu,Heidenreich:2020pkc}. 

For the purposes of computing torsion in cohomology,  to consider these new terms in the Lagrangian may seem redundant, since in the type IIA constructions that we have discussed they are related to the same kind of torsion groups, namely ${\rm Tor} H_3(X_6, \IZ) \simeq$ ${\rm Tor} H_2(X_6, \IZ)$ and their linking number. However, an important difference is that the terms \eqref{p=2} and \eqref{p=3} appear in 4d $\CN=1$ string theory vacua, like in type II orientifold compactifications, while \eqref{stuck} and \eqref{stuckdual} typically appear in 4d $\CN=2$ compactifications without vacua, like the example considered in \cite{Gurrieri:2002wz}.  In this case the $\CN=2$ supersymmetry of the Lagrangian is realised off-shell, while solutions to the equations of motion at most preserve a fraction of this supersymmetry, like the domain-wall solution preserving four supercharges to be discussed in the next section. In practice this implies that the 10d background is not of the form $\IR^{1,3} \times X_6$, but instead a fibration of $X_6$ over a real line or a plane in $\IR^{1,3}$. Following the general philosophy of \cite{Gurrieri:2002wz,Grana:2005ny} we are entitled to carry out the usual procedure of dimensional reduction to 4d -- and therefore our discussion above -- as long as the variation of this fibration is very small compared to the compactification scale. The only additional thing that we need to take into account is that for Conjecture \ref{conj:BPS} to apply the objects like 4d strings and particles must be BPS with respect to the 4d solution, which is a stronger condition that being BPS in a would-be $\CN=2$ vacuum. In practice, this means that they must be calibrated also from the point of view of the fibration, as we will illustrate in the next section. 

The fact that AB strings and particles cannot be BPS in $\CN=1$ orientifold compactifications seems to clash with the proposal in \cite{Camara:2011jg}, in the sense that the basis of non-harmonic $p$-forms in which one expands the 10d RR potentials to obtain a St\"uckelberg Lagrangian \eqref{stuck} cannot come from smearing the delta-forms of calibrated cycles. Nevertheless, one can still make sense of such a basis of non-harmonic forms if one considers the extension of Conjecture \ref{conj:BPS} formulated around \eqref{deldif}. For instance, one could try to describe the torsion two- and three-cycles of a Calabi--Yau threefold as the difference of two calibrated cycles with equal volume, or some other combination of calibrated cycles. As long as there are some eigenmodes below the compactification scale that couple differently to these calibrated cycles, there will be non-harmonic $p$-forms that one builds from smearing their bump delta-forms. Finally, one should make sure that such harmonic forms have the appropriate parity under the orientifold action to lead to a St\"uckelberg term.

%%%%%%%%%%%%%%%%%%%
%%%%%%%%%%%%%%%%%%%

\section{A simple example}
\label{s:simple}

The simplest example of SU(3)-structure manifolds with torsion in cohomology are nilmanifolds or twisted tori, which in the context of type II string compactifications were initially considered in \cite{Gurrieri:2002wz,LopesCardoso:2002vpf,Kachru:2002sk}. Particularly interesting for our discussion is the setup of \cite{Gurrieri:2002wz}, in which the simplest kind of twisted torus is realised as a 4d domain-wall solution. In the following we will see how the objects defined in the previous sections, in particular torsion calibrated cycles and their smeared delta sources, are described in this case. 

\subsection{The 10d background}

Let us recall the main idea behind the construction in \cite{Gurrieri:2002wz}. One first considers a toroidal compactification of type IIB string theory to 4d with a backreacted NS5-brane wrapping a special Lagrangian three-cycle of ${\bf T}^6$ and extended along $\IR^{1,2} \subset \IR^{1,3}$ in the non-compact dimensions. The long-wavelength approximation  of this backreaction provides a domain-wall solution in 4d, which upon three T-dualities in ${\bf T}^6$ becomes a type IIA background with constant dilaton and a twisted six-torus $\tilde{\bf T}^6$ fibered over a non-compact direction. A simple generalisation of this setup results in the following type IIA 10d string frame background:
\bea
\label{SU3ex}
ds^2 &= &ds^2_{\IR^{1,2}}+ \ell_s^2 V(d\xi)^2 + \ell_s^2 ds^2_{\tilde{\bf T}^6} , \\ 
ds^2_{\tilde{\bf T}^6}& = &(2\pi)^2 \left[\frac{R_1^2}{V_1}(\eta^1)^2+\frac{R_2^2}{V_2}(\eta^2)^2+\frac{R_3^2}{V_3}(\eta^3)^2+\frac{V R_4^2}{V_1}(\eta^4)^2+\frac{V R_5^2}{V_2}(\eta^5)^2+\frac{V R_6^2}{V_3}(\eta^6)^2 \right] ,
\label{SU3exb}
\eea
where $\xi$ is the 4d coordinate transverse to the domain-wall, $R_i$ are radii measured in string units, and $\eta^i$ are the left-invariant one-forms of the twisted six-torus, defined as
\begin{equation}
\begin{array}{ll}
\eta^1= d x^1 + M_1  x^6 d x^5   \,, & \qquad \eta^4= d x^4 \,, \\
\eta^2= dx^2 + M_2 x^4 d x^6  \,, & \qquad \eta^5= d x^5 \,, \\
\eta^3= d x^3 + M_3 x^5 d x^4   \,, & \qquad \eta^6=  d x^6 \, ,
\end{array}
\label{etas}
\end{equation}
with $M_i \in \mathbb{N}$. Finally, 
\be
V = V_1V_2V_3 , \qquad V_i = 1 - \zeta_i \xi , \qquad \zeta_i = \frac{M_i}{2\pi} \frac{R_i R_{i+3}}{R_4R_5R_6} .
\label{Vis}
\ee
To recover the case of \cite{Gurrieri:2002wz} one needs to take $M_i=M \in \mathbb{N}$ and $M_j= M_k=0$, with $i \neq j \neq k \neq i$. The solution applies to the range $(0,\xi_{\rm end})$, with $\xi_{\rm end} = {\rm min} \{\zeta_i^{-1}\}_i$, while for $\xi<0$ one should glue a direct product $\IR^{1,3} \times {\bf T}^6$, with torus radii $R_i$.\footnote{Our background differs slightly from the one  in \cite{Gurrieri:2002wz}, in the sense that therein the choice $V_i = \zeta_i \xi$ along the range $\xi \geq 0$ is taken, for a domain wall  at $\xi=0$. Both choices are compatible with the domain-wall analysis of \cite{Curio:2000sc,Behrndt:2001qa,Behrndt:2001mx}, but we find that our choice also reproduces the scalar flow features of $\oh$BPS domain walls in $\CN=1$ EFTs (see e.g.  \cite[section 4.3.2]{Lanza:2020qmt}) and is compatible with the presence of BPS AB strings as particles, as discussed below. \label{ft:dw}}

We refer to \cite{Marchesano:2006ns} for more details on the geometry and topology of this class of twisted six-tori. As in there, one can impose a $\IZ_2 \times \IZ_2$ orbifold projection that reduces the structure of the internal manifold to a genuine SU(3) structure, and which we will assume in the following. In the conventions $d{\rm vol}_{X_6} = -\frac{1}{6} J^3 = \frac{i}{8} \bar{\Omega} \wedge \Omega$, the SU(3)-structure calibrations $(J,\Omega)$ are given by
\bea
\label{Jex}
J & = & 4\pi^2 \left( t^1 \,\eta^1 \wedge \eta^4 +t^2\, \eta^2 \wedge \eta^5  + t^3 \,\eta^3 \wedge \eta^6 \right) , \\
\Om & = & {i} (2\pi)^3 V^{-1/2} R_1R_2R_3 \left( \eta^1 + i  \tau^1 \eta^4 \right) \wedge \left( \eta^2 + i \tau^2 \eta^5 \right) \left( \eta^3 + i \tau^3 \eta^6 \right) ,
\label{Omex}
\eea
with
\be
\label{defttau}
t^i = \frac{V^{1/2}}{V_i} R_iR_{i+3} , \qquad  \tau^i = V^{1/2} \frac{R_{i+3}}{R_i} .
\ee

The calibrated objects of this SU(3)-structure manifold are those $p$-chains whose volume is computed by integrating $\Omega$ or $e^{iJ}$. Recall, however, that we are interested in a particular kind of calibrated cycles. First, they need to be calibrated in a strict sense, meaning that upon varying the values of the $R_i$ they are still calibrated. Second, they need to be mutually BPS with the domain-wall source, in order to be actual BPS objects of the background \eqref{SU3ex}. This second criterion is more easily analysed in the type IIB mirror background, as done in Appendix \ref{ap:NS5DW}. In our context, one finds the following BPS objects that are relevant to our discussion:

\begin{itemize}

\item[-] A D4-brane wrapped on $\Pi_3^{\rm tor} = \{ x^4=x^5=x^6=0\}$ in $\tilde{\bf T}_6$ and extended along $\xi$.  

\item[-] A D4-brane wrapped on $\Sigma_4^i = \{ x^i = x^{i+3} = 0\}$ in $\tilde{\bf T}_6$.

\item[-] An Euclidean D2-brane on $\Pi_2^i = \Sigma_4^j \cap \Sigma_4^k$, with $i \neq j \neq k \neq i$ in $\tilde{\bf T}_6$ and extended along $\xi$.

\item[-] An Euclidean D2-brane wrapped on $\Sigma_3 = \{ x^1=x^2=x^3=0\}$ in $\tilde{\bf T}_6$.

\end{itemize}
Notice that, when extending a D-brane along $\xi$, it does not make sense to do it beyond  $\xi_{\rm end}$, where the metric degenerates and we enter a strong coupling region. 

The submanifolds $(\Pi_3^{\rm tor}, \Sigma_4^i, \Pi_2^i, \Sigma_3)$ and others can be described via group theory techniques, by first writing the twisted six-torus as a coset $\tilde{\bf T}^6 = G/\Gamma$, with $G$ a Lie group of a 2-step nilpotent algebra and $\Gamma$ a co-compact lattice, and then exponentiating different set of generators of $G$, see \cite[Appendix A]{Marchesano:2006ns}. Using this framework and the results of \cite{Nomizu1954OnTC,Cenkl2000NILMANIFOLDSAA}, one can see that 
\be
{\rm Tor } H_3 (\tilde{\bf T}^6, \IZ)_{\IZ_2 \times \IZ_2} =  {\rm Tor} H_2 (\tilde{\bf T}^6, \IZ)_{\IZ_2 \times \IZ_2} = \IZ_M ,
\label{cohott6}
\ee
where $M = {\rm g.c.d.} (M_1, M_2, M_3)$, and the subindex represents those cycles invariant under the $\IZ_2 \times \IZ_2$ orbifold projection.\footnote{The generators of the $\IZ_2 \times \IZ_2$ orbifold group act on the left-invariant one forms as $\theta_1: (\eta^1, \eta^2, \eta^3, \eta^4, \eta^5, \eta^6) \mapsto  (\eta^1, -\eta^2, -\eta^3, \eta^4, -\eta^5, -\eta^6)$ and $\theta_2: (\eta^1, \eta^2, \eta^3, \eta^4, \eta^5, \eta^6) \mapsto  (-\eta^1, -\eta^2, -\eta^3, -\eta^4, -\eta^5, \eta^6)$ \cite{Marchesano:2006ns}. It is not obvious if the torsion cohomology of the orbifold quotient $\tilde{\bf T}^6/\IZ_2 \times \IZ_2$ corresponds to \eqref{cohott6} or if it has further elements. However, in case that some additional torsion cycles existed, one can show that they are not calibrated and they do not couple to any light eigenmode. Therefore one can ignore them for the purposes of this work. \label{ft:orbifold}}
The three-cycle $\Pi_3^{\rm tor}$ is the generator of ${\rm Tor } H_3 (\tilde{\bf T}^6, \IZ)_{\IZ_2 \times \IZ_2}$, while ${\rm Tor } H_2 (\tilde{\bf T}^6, \IZ)_{\IZ_2 \times \IZ_2}$ is generated by $\Pi_2^{\rm tor} = \sum_i (M_i/M) \Pi_2^i$. Additionally, $\Sigma_3$ is a three-chain with a boundary homotopic to $M \Pi_2^{\rm tor}$ and, if $M_i \neq 0$, $\Sigma_4^i$ is a four-chain with a boundary homotopic to $M_i \Pi_3^{\rm tor}$. All these $p$-chains are calibrated by either $\Omega$ or $e^{iJ}$, with a calibration phase that will depend on their orientation. The D-branes listed above are $\oh$BPS in the background \eqref{SU3ex}, which means that they preserve two supercharges out of the four supercharges preserved by the solution. The two supercharges that they preserve will depend on their orientation. For instance, a D4-brane wrapping $\Sigma_4^i$ looks like a $\oh$BPS monopole in 4d, and preserves two supercharges of the domain-wall solution. Reversing the orientation and wrapping the D4-brane on $-\Sigma_4^i$ corresponds to a 4d $\oh$BPS monopole with opposite charge and preserving the other two supercharges of the background, an object that we will refer to as anti-BPS monopole. Here we will not keep track of which objects preserve which supercharges, because a much more straightforward picture will arise when we interpret this system in terms of Hitchin flow equations. 

A D4-brane wrapping a chain $\Sigma_4^i$ with a boundary is not consistent by itself, as it develops a worldvolume anomaly, but one can make it consistent by attaching D4-branes wrapped on $\p \Sigma_4^i$. In the present setup, if the D4-brane wrapping $\Sigma_4^i$ is located at $\xi_0 \in (0, \xi_{\rm end})$, one can cure its worldvolume anomaly by wrapping $M_i$ D4-branes on $\Pi_3^{\rm tor}$, and connecting them to $\p \Sigma_4^i$. These $M_i$ D4-branes will look like 4d strings that extend along the coordinate $\xi$, and either end on an anti-monopole in a different location, or stretch up until the origin $\xi=0$. From the 4d perspective, in the first case we have a monopole-anti-monopole pair connected by $M_i$ AB strings, as expected for a 4d EFT with a Lagrangian of the form \eqref{stuckdual} and a monopole of charge $M_i/M$. In the second case, we have a 4d avatar of a Hanany-Witten brane creating effect \cite{Hanany:1996ie}, mirror dual to a D3-brane crossing the NS5-brane (the domain wall), with $M_i$ D1-branes stretching along both after the crossing. As stressed in \cite{Berasaluce-Gonzalez:2012awn}, this effects also signal the presence of a discrete gauge symmetry, encoded either in the Lagrangian \eqref{stuckdual} or its dual. Similarly, the worldvolume anomaly of an Euclidean D2-brane in $\Sigma_3$ can be cured by $M$ D2-branes wrapped on $\Pi_2^{\rm tor}$ and connected to $\p \Sigma_3$. From the 4d viewpoint this is perceived like $M$ AB Euclidean particles ending on an instanton \cite{Banks:2010zn}. Notice that in this case a configuration made of $\oh$BPS objects involves $M$ Euclidean AB particles extended along the coordinate $\xi$, that stretch either between the domain-wall source and the instanton or between an instanton-anti-instanton pair.

In terms of these 4d objects one can compute the quantities $f$ and $g$ that feature the discussion of section \ref{s:dimred}, by using \eqref{fst} and \eqref{gpt}. Since in our example there are many axions and gauge bosons, in order to isolate a pair of them in the Lagrangian, as in \eqref{stuck}, we must consider the particular case $M_i \neq 0$, while $M_j = M_k =0$ for $i\neq j \neq k \neq i$. One then finds 
\be
f = \left( \tau^1\tau^2 \tau^3 \right)^{-1/2} , \qquad g = \frac{1}{2\pi} \sqrt{\frac{t^i} {t^j t^k }} .
\label{fandg}
\ee
Additionally, via the direct computation of section \ref{s:direct} (see eq.\eqref{eq: U0T3}) or the results of Appendix \ref{ap:spectra}, one obtains that the smallest non-vanishing eigenvalue of $\tilde{\bf T}^6$ is
\be
\lambda_{\rm st} =  \zeta_i V_i^{-3/2} ,
\label{lamstex}
\ee
and so it follows that the first equality in \eqref{Lphys} is satisfied with $N=M_i$, even if all quantities depend on the coordinate $\xi$. %Notice that the quotient $m_{\rm KK}/m_{\rm st}$ (where $m_{\rm KK}$ is set by the largest radius), decreases with $V$ as one moves away from the domain wall source at $\xi=0$. 

One can also see that the unit-norm exact three-form eigenmode corresponding to \eqref{lamstex} is
\be
b_3 = f^{-1}  \eta^4 \wedge \eta^5 \wedge \eta^6  = f^{-1} \delta_3^{\rm sm} (\Pi_3^{\rm tor}),
\ee
where in the second equality we have again used the results of section \ref{s:direct}, cf. eq.\eqref{deltasmT6}. We thus find perfect agreement with the discussion of section \ref{s:dimred}, in which the definition of $f$ via a smeared delta bump-form coincides with the value in \eqref{fst}. A similar check can be made for $g$, and the combined result is such that Conjecture \ref{conj:BPS} is verified. In the following we will discuss how to extend this result to general $M_i \in \mathbb{N}$, using the 4d EFT description. 

\subsection{EFT description}
\label{ss:EFTdesc}

To obtain the 4d effective description of this system one may follow the approach in \cite{Gurrieri:2002wz}, or its extension to more general setups discussed in \cite{Grana:2005ny,Kashani-Poor:2006ofe}. One first defines the following basis of three-forms 
\begin{subequations}
\label{alphabetas}
\begin{align}
   \a_0 = \eta^1 \wedge \eta^2 \wedge \eta^3 , & \qquad \beta^0 = \eta^4 \wedge \eta^5 \wedge \eta^6 , \\
   \a_1 = \eta^4 \wedge \eta^2 \wedge \eta^3  , & \qquad \beta^1 = - \eta^1 \wedge \eta^5 \wedge \eta^6 , \\
   \a_2 = \eta^1 \wedge \eta^5 \wedge \eta^3  , & \qquad \beta^2 =  - \eta^4 \wedge \eta^2 \wedge \eta^6 , \\
   \a_3 = \eta^1 \wedge \eta^2 \wedge \eta^6  , & \qquad \beta^3 = - \eta^4 \wedge \eta^5 \wedge \eta^3 ,   
\end{align}
\end{subequations}
and a basis of two- and four-forms
\begin{subequations}
\label{omegas}
\begin{align}
\omega_1 = \eta^1 \wedge \eta^4 , \qquad \omega_2 = \eta^2 \wedge \eta^5 , \qquad \omega_3 = \eta^3 \wedge \eta^6 , \\
\tilde{\omega}^1 = - \omega_2 \wedge \omega_3, \qquad \tilde{\omega}^2 = - \omega_3 \wedge \omega_1, \qquad  \tilde{\omega}^3 = - \omega_1 \wedge \omega_2 .
\end{align}
\end{subequations}
This set of forms are those that are invariant under the $\IZ_2 \times \IZ_2$ projection that takes us to a genuine SU(3)-structure. Notice that they satisfy $\int_{\tilde{\bf T}^6} \a_i \wedge \beta^j = \int_{\tilde{\bf T}^6} \om_i \wedge \tilde{\om}^j = \delta_i^j$, with a specific normalisation which is crucial for the discussion that follows, since we are going to expand both the calibrations $(J, \Omega)$ and the 10d RR fields in these forms, and the latter are going to define the 4d axion periodicities the global $U(1)$ gauge transformations. While for harmonic $p$-forms one has a clear prescription to define an integral basis, the same is not true for exact and co-exact elements of this set, which can be identified thanks to the relations
\be
d\om_i = -M_i \b^0 , \qquad d\a_0 = -M_i \tilde{\om}^i .
\ee
In the present setup such a normalisation can be fixed by means of mirror symmetry, just as in \cite{Gurrieri:2002wz}. However, in the general setting of \cite{Grana:2005ny,Kashani-Poor:2006ofe} it is simply assumed as an input. In section \ref{s:general} we will argue that one can fix it by defining the set $\{\a_A, \b^B, \om_a, \tilde{\om}^b\}$ as smeared delta-forms. 

Following \cite{Grana:2005ny,Kashani-Poor:2006ofe} we expand the NS-NS sector in the above basis
\bea
\label{JcandOm}
J_c  &= &B+ iJ = 4\pi^2 (b^j + i t^j) \, \om_j , \\ \nonumber
\Om  &=  & {i} (2\pi)^3 \sqrt{\frac{t^1t^2t^3}{\tau^1\tau^2\tau^3}} \left( \a_0 + z^i \a_i - z^2 z^3 \b^1 - z^1 z^3 \b^2 - z^1 z^2 \b^3  + z^1z^2z^3 \b^0 \right), \quad  z^j = a^j + i \tau^j .
\eea
Similarly, one expands the RR potential $C_3$ as
\be
 C_3 = 2\pi \ell_s^{3} \left[ A_1^i \wedge \om_i +  \theta^I \alpha_I + \tilde{\theta}_K \beta^K \right] ,
\ee
where $(\theta^I, \tilde{\theta}_I)$ with $I= (0,i)$ represent axions of unit periodicity. The dimensional reduction of this term gives 
\be
(2\pi)^2 \left[g_{ii} (dA_1^i)^2 + \hat{f}^{00} \left(d\tilde{\theta}_0 - M_i A_1^i\right)^2  + \hat{f}^{ii} (d\tilde{\theta}_i)^2 + \hat{f}_{II} (d{\theta}^I)^2  \right] ,
\label{StuckN=2gen}
\ee
plus a mass term for $\theta^0$. Here we have defined
\be
g_{ii} = (2\pi)^2\, \frac{t^j t^k}{t^i} , \qquad \hat{f}^{00} = e^{2\phi_4} (\tau^1\tau^2\tau^3)^{-1} M_{\rm P}^2 , \qquad \hat{f}^{ii} = e^{2\phi_4} \frac{\tau^i}{\tau^j\tau^k}  M_{\rm P}^2 ,
\ee
with $i \neq j \neq k \neq i$, and $\hat{f}_{II} = (\hat{f}^{II})^{-1} e^{4\phi_4}M_{\rm P}^4$, where $e^{\phi_4} = e^{\phi}/\sqrt{4\pi\,t^1t^2t^3}$ is the 4d dilaton. Notice that all these couplings depend on the domain-wall transverse coordinate $\xi$, while the axion vevs remain constant along it. The NS-NS sector of the compactification varies along $\xi$ via the non-trivial profile of the saxions $t^i, \tau^i$ along this coordinate, as captured by \eqref{defttau}, and in agreement with the results of \cite{Behrndt:2001mx,Mayer:2004sd}.

The lightest massive $p$-form mode has the following squared mass
\be
m_{\rm st}^2 = V^{-1} \left[\sum_i \frac{\zeta_i^2}{V_i^2} \right] e^{2\phi_4} M_{\rm P}^2 ,
\label{lamstexgen}
\ee
and so it is a priori not obvious how to compute the smeared linking number using \eqref{Lphys}. To do so, one must take into account that for generic $M_i$'s the torsion two-cycle $\Pi_2^{\rm tor} = \sum_i (M_i/M) \Pi_2^i$ is not a smooth calibrated cycle, but instead a linear combination of them. In this case, it is the extension of the conjecture made around \eqref{deldif} that should be applied. One obtains 
\be
\delta^{\rm sm}_4 (\Pi_2^{\rm tor}) =  \sum_i \frac{M_i}{M} \tilde{\om}^i ,
\ee
whose projection into the four-form eigenmode with the smallest non-vanishing eigenvalue gives
\be
g =   \frac{1}{2\pi}  \sqrt{ \sum_i \frac{M_i^{2}}{M^2} \frac{t^i} {t^j t^k }} .
\label{ggenM}
\ee
Defining $f = \sqrt{\hat{f}^{00}}$ one reproduces \eqref{Lphys} with $N=M = {\rm g.c.d.} (M_1, M_2, M_3)$, as expected. From a purely 4d EFT viewpoint, one can interpret \eqref{ggenM} as the gauge coupling of the linear combination of $U(1)$'s that develops a St\"uckelberg mass.

\subsection{Hitchin flow equations}
\label{ss:hitchin}

As already pointed out in \cite{Gurrieri:2002wz}, the background \eqref{SU3ex} can be understood geometrically as a fibration of a half-flat manifold $X_6$ over a real coordinate, that gives a seven-dimensional $G_2$-manifold $Y_7$. The general description of this kind of fibrations has been given in \cite{hitchin2001stable,chiossi2002intrinsic}, and are known as Hitchin flow equations. In the standard description, the real coordinate $z$ has a flat metric, and one constructs the $G_2$-structure forms
\bea
\varphi & = & dz \wedge J  - {\re \Om} , \\
*\varphi & = & -dz \wedge {\im \Om} - \oh J \wedge J ,
\eea
where $J$ and $\Om$ are the $z$-dependent SU(3)-structure calibrations of $X_6$. Demanding that  $Y_7$ has $G_2$ holonomy amounts to impose that $\varphi$ is harmonic in $Y_7$. If we describe the 7d derivative as 
\be
d_7 = \p_z  dz \wedge  + \, d ,
\ee 
with $d$ the exterior derivative along the 6d fibre, this requirement reads
\bea
\label{dreom}
d {\im \Om} & = & \oh\p_z \left(J \wedge J\right) , \\
d J & = & - \p_z {\re \Om} .
\label{dj}
\eea
 In our background the coordinate $\xi$ has a non-trivial metric, more precisely $dz = - V^{1/2}d\xi$, where the sign choice accounts for the difference in our background compared to \cite{Gurrieri:2002wz} (see footnote \ref{ft:dw}). The Hitchin flow equations then take the following form :
\bea
\label{dreomxi}
V^{1/2} d {\im \Om} & = & - \oh \p_\xi \left(J \wedge J\right) , \\
V^{1/2} d J & = &  \p_\xi {\re \Om} . 
\label{djxi}
\eea
Applied to the background \eqref{SU3ex} these equations reduce to
\be
\p_\xi V_i  =  - \zeta_i ,
\ee
which is clearly satisfied by \eqref{Vis}.

The Hitchin flow equations have a nice interpretation when it comes to D-branes on torsion cycles, that can be illustrated explicitly in the solution \eqref{SU3ex}. Let us consider $M$ D4-branes wrapped on $\Pi_3^{\rm tor}$ and extended along an interval $(0, \xi_0) \subset (0,\xi_{\rm end})$. At $\xi_0$  one places a D4-brane wrapping a four-chain $\Sigma_4$, such that its boundary coincides with the $M$ torsion three-cycles. From the 4d viewpoint, this represents a 4d monopole in which $M$ AB strings end, with their other end at the domain-wall source. Since both sets of D4-branes yielding the monopole and the AB strings are calibrated by $*\varphi$, they must be mutually BPS, and satisfy the marginal stability condition \eqref{BPSmono}. Additionally, the total energy of the system must be given by its central charge, which is the integral of $*\varphi$ over the full D4-brane worldvolume in the $G_2$ manifold $Y_7$, and it is easy to argue that this central charge must be independent of the monopole position $\xi_0$. 

Indeed, notice that shifting the value of $\xi_0$ corresponds to add $M$ AB strings extended along the interval $(\xi_0, \xi_0') \subset (0,\xi_{\rm end})$, with a D4-branes wrapping $\Sigma_4$ at each end, with opposite orientations. From the 4d viewpoint, this realises a monopole-anti-monopole pair in which AB strings end. Since this object can annihilate by itself, one expects that its central charge vanishes. Microscopically, the whole object corresponds to a trivial four-cycle in the $G_2$ manifold $Y_7$, and so since $*\varphi$ is closed its integral must vanish on it. So indeed the monopole-anti-monopole pair carries no central charge and changing the value of $\xi_0$ in the above BPS configuration should not change the energy of the system. In particular this energy should match that of a monopole placed at $\xi = \xi_{\rm end}$ which is equivalent to having $M$ AB strings, and to a monopole at $\xi =0$, which does not have any AB strings attached to it. 

This is indeed what the Hitchin flow equations are telling us, and in particular \eqref{dreom}. On the one hand, $\p_z J \wedge J$ represents the variation of the mass of BPS monopoles  when we move along $z$. On the other hand, ${\im \Omega}$ integrated along the torsion three-cycle measures the tension of a BPS 4d AB string, and by Stokes' theorem, this is equivalent to integrating ${d \im \Omega/M}$ over the four-chain $\Sigma_4$ linking $M$ of them. So what \eqref{dreom} is saying is that it is the monopoles in which $M$ AB strings can end the ones whose mass varies along the coordinate transverse to the domain wall. Moreover, there is a mass scale associated to the 4d string, which is its tension integrated along the interval $(0, \xi_0)$. For BPS objects, this energy increases with $\xi$ at the same rate as the monopole mass decreases, and that is why the total central charge and therefore the energy of the system stays constant. In our example \eqref{SU3ex} one can see that the factors of $V$ cancel for a 4d AB string, so the energy of $M_i$ BPS AB strings is given by $\ell_s^{-1} {\rm Vol}(\Pi_3^{\rm tor}) M_i \xi_0 \propto \xi_0 $. Additionally, the mass of the monopole in which such strings can end is given by $\ell_s^{-1} {\rm Vol}(\Sigma_4^i) \propto V_i|_{\xi_0}$. Therefore, it decreases linearly with $\xi_0$, precisely compensating the change in the energy of the AB strings.\footnote{One can engineer the BPS configuration of $M$ AB strings ending on a monopole by a Hanany-Witten brane-creation effect, as one can check using the mirror type IIB picture, see Appendix \ref{ap:NS5DW}. The interpretation is then that a Hanany-Witten effect does not change the energy of a BPS object. The mass of a monopole located at $\xi_0 \in (-\infty, 0)$ and at $\xi_0 \in (0,\xi_{\rm end})$ is the same, if in the second case we include the energy of the extended AB strings. That is, if at both sides we compute the energy or central charge of the gauge invariant operator.}

This example illustrates how \eqref{BPSmono} can be satisfied, and the expectation of subsection \ref{ss:BPS}, that one should be able to add up central charges of BPS objects in $\IZ$, even when their topological charge is $\IZ_N$. In the case at hand, $M$ D4-branes wrapping $\Pi_3^{\rm tor}$ can disappear by ending on a monopole, but a monopole nucleation process costs  energy, which is minimised for the case of BPS monopoles. The discussion above implies that this energy is at least that of $M$ BPS strings extended along the interval $(\xi_0, \xi_{\rm end})$. Therefore nucleating a monopole at $\xi_0$ is topologically possible, but not energetically favoured. In this sense, adding up an arbitrary number of  AB strings is well-defined in the BPS context, as well as considering a cone of 4d AB string charges.

%%%%%%%%%%%%%%%%%%%
%%%%%%%%%%%%%%%%%%%

\section{Direct computation}
\label{s:direct}

While the general arguments of section \ref{s:dimred} motivate Conjecture \ref{conj:BPS} for torsion cycles in SU(3)-structure manifolds, it is instructive to work out in detail how the conjecture is realised in explicit examples, by a direct comparison of the torsion linking number and its smeared version. In this section we perform such a comparison for the SU(3)-structure manifold of section \ref{s:simple}, more precisely for the twisted six-torus with a single metric flux. As we will see, the direct computation of the torsion linking number displays a series of cancellations between terms that is reminiscent of those that occur in the computation of topological indices, and that leaves the smeared torsion linking number \eqref{Lsm} as the only non-vanishing contribution. The reader not interested in these technical details may safely skip to the next section.

\subsubsection*{The setup}

Let us consider the twisted six-torus background in \eqref{SU3exb}, rewritten as
\be
ds^2_{\tilde{\bf T}^6} = (2\pi)^2 \sum_i \left(\frac{t^i}{\tau^i}(\eta^i)^2+t^i\tau^i(\eta^{i+3})^2 \right) ,
\label{SU3exc}
\ee
and with the definitions \eqref{defttau} and \eqref{etas}. In particular we consider $M_i \neq 0$ and $M_j = M_k =0$ with $i \neq j\neq k \neq i$. In this case, the metric background factorises as $\tilde{\bf T}^6 = \tilde{\bf T}^3 \times {\bf T}^3$, and all the torsion cycles correspond to a direct product of a torsion one-cycle in $\tilde{\bf T}^3 \simeq \langle x^i, x^{j+3}, x^{k+3} \rangle$ and a non-trivial cycle in ${\bf T}^3 \simeq \langle x^j, x^k, x^{i+3}\rangle$. As a result, all torsion linking numbers of $\tilde{\bf T}^6$ stem from the torsion linking numbers between one-cycles in $\tilde{\bf T}^3$. Moreover, the calibration condition in $\tilde{\bf T}^6/\IZ_2 \times \IZ_2$ will translate into a subset of such torsion one-cycles. Therefore our strategy will be to verify Conjecture \ref{conj:BPS} for such a subset, then extend the result into calibrated two and three-cycles of $\tilde{\bf T}^6$, and finally check that the $\IZ_2 \times \IZ_2$ projection does not modify the statement. A necessary first step is to describe the set of massive $p$-form modes in $\tilde{\bf T}^3$, which one can accomplish using a general method for three-manifolds with isometries. 

\subsubsection*{Massive spectra of three-manifolds}

To describe the massive $p$-form spectrum of a twisted three-torus, one may use the method of  \cite{Ben_Achour_2016}, which applies to compact Riemannian three-dimensional manifolds $X_3$ with a continuous isometry.  Such a manifold admits a unit-norm Killing vector $\chi$, and we assume that its dual one-form satisfies 
\begin{equation}
	\star d \chi = \lambda_\chi \,\chi \,, \qquad \Delta_3 \chi = \lambda_\chi^2\, \chi \,, \qquad \chi^2 = 1  \,, \qquad  \lambda_\chi, \in \IR \, ,
\end{equation}
and that its integral curves are closed. Here $\star$ and $\Delta_3$ stand for the Hodge star operator and the Laplacian on $X_3$, respectively. Then, let $\{\phi_\a\}$ be an orthonormal basis of complex scalar eigenforms of the Laplacian such that\footnote{Notice that such 
 a basis always exists because $[\Delta_3, \mathcal{L}_\chi ] = 0$. }
\begin{equation} \label{eqbasisscalars}
	\Delta_3 \phi_\a = \sigma_\a^2 \phi_\a \,,\qquad \mathcal{L}_\chi \phi_\a = i \mu_\a \phi_\a \,, \qquad   \sigma_\a \in \mathbb{R} \,, \quad  \mu_\a \in \mathbb{R}\,.
\end{equation} 
Solving the second condition of (\ref{eqbasisscalars}) we can obtain the explicit dependence of  the $\{\phi_\a\}$ on the isometry coordinate $\th$ associated to $\chi$
\begin{equation}\label{eqphiexp}
	\phi_\a = e^{i \mu_\a \theta} K_\a \,, \qquad \th \sim \th + 2\pi r \, ,
\end{equation}
with $d\theta =\chi$ and $K_\a$ functions which do not depend on $\th$. Given that $\th$ parameterises a closed integral curve of radius $r$ in a compact manifold, we obtain the quantisation condition $\mu_\a r \in  \mathbb{Z} $\,. 

In this setup, it is possible to give a simple description of non-harmonic eigen-one-forms of the Laplacian, in terms of the Killing vector $\chi$ and the scalar eigenforms $\phi_\a$. We define 
\begin{equation}
	R_\a = d \phi_\a \,, \qquad S_\a = \star d (\phi_\a \chi ) \,, \qquad T_\a = \star d S_\a \,. \label{eq: BandC}
\end{equation}
It is easy to see that the set $R_\a$ forms a complete basis of exact eigen-one-forms. The set of co-exact one-forms $S_\a$ and $T_\a$ is closed under the action of the operator $ \star d$
\begin{equation}
	\star d S_\a = T_\a \,, \qquad \star d T_\a = \sigma_\a^2 S_\a +  \lambda_\chi T_\a\, ,
\end{equation}
from where one can find the following eigenforms of $\star d$
\begin{equation}
	U^\pm_\a = \left(\frac{1}{2}\pm\frac{ \lambda_\chi}{2 \sqrt{ \lambda_\chi^2 + 4 \sigma_\a^2}}\right) T_\a \pm \frac{\sigma_\a^2}{\sqrt{ \lambda_\chi^2 + 4 \sigma_\a^2}}S_\a \,.  \label{eq: Dpm}
\end{equation}
Therefore, since the action of the Laplacian $\Delta_3$ on co-closed forms amounts to $\star \, d \star d$, we obtain that the $U^\pm$ are eigenforms of the Laplace operator with eigenvalues
\begin{equation}
	(\lambda^\pm_\a)^2 = \sigma_\a^2 + \frac{ \lambda_\chi^2}{2} \pm \frac{ \lambda_\chi}{2}\sqrt{ \lambda_\chi^2 + 4 \sigma_\a^2 } \,.\label{eq: lambdapm}
\end{equation} 
Let us dub the constant eigenmode of the Laplacian as $\phi_0 = 1/\sqrt{V_3}$, with $V_3$ the volume of $X_3$. Then the eigenmode $U_0^-$ identically vanishes, while $U_0 \equiv U_0^+$ has eigenvalue $ \lambda_\chi^2$  with respect to $\Delta_3$ and takes form $U_0 =  \lambda_\chi^2 \, \chi \phi_0$. Moreover, the set of co-exact one-forms $U^\pm_\a $ are normalised to unity by multiplying them by the following factor 
\begin{equation}
c_\a^{\pm} = \left[\frac{( \lambda_\chi^4+3 \lambda_\chi^2 \sigma_\a^2+\sigma_\a^4) - ( \lambda_\chi^2+\sigma_\a^2)\,\mu_\a^2}{2} \pm   \lambda_\chi \, \frac{( \lambda_\chi^4 + 5 \lambda_\chi^2 \sigma_\a^2 + 5\sigma_\a^4)-( \lambda_\chi^2+3\sigma_\a^2)\,\mu_\a^2}{2 \sqrt{ \lambda_\chi^2 + 4 \sigma_\a^2}}\right]^{-1/2}\,.
\end{equation}
In the following we will assume that the $U_\a^{\pm}$ have been normalised to unit norm. In particular, we have that $c_0 \equiv c_0^+ =  \lambda_\chi^{-2}$, and so $U_0 = \chi \phi_0$. 

The set $\{U_\a^\pm\}$ is part of the co-exact one-form eigenspectrum of $X_3$, but the above method does not guarantee that it is a complete set. In the particular case of $\tilde{\bf T}^3$ one can check that the whole co-exact spectrum is of this form, as verified in Appendix \ref{ap:spectra} by using the results of  \cite{Andriot:2018tmb}.

\subsubsection*{Computing the linking number}

Using that $\{U_0, U_\a^\pm\}$ is a complete basis of co-exact eigen-one-forms of $X_3$,\footnote{There may be more than one eigenform for a given eigenvalue, but this will not change our final result.} we can expand the bump delta two-form for a torsion one-cycle $\pi_1 \subset X_3$ as
\begin{equation}
	\delta^{(2)}(\pi_1) = K_0 \star U_0 + \sum_\a \left( K_\a^+  \star U^+_\a + K_\a^-  \star U^-_\a \right)\, , \label{delta2sm}
\end{equation} 
where
\be
K_0 = \int_{\pi_1} U_0, \qquad K_\a^\pm = \int_{\pi_1} U_\a^\pm \, .
\ee
In terms of these expressions, the linking number \eqref{L} between two torsion one-cycles reads
\begin{equation}
	L(\pi, \tilde{\pi}) = \frac{1}{ \lambda_\chi} K_0 \tilde{K}_0 + \sum_\a \left[ \frac{1}{\lambda_\a^+} K^+_\a \tilde{K}^+_\a + \frac{1}{\lambda_\a^-} K^-_\a \tilde{K}^-_\a \right]\, ,
	\label{LtT3}
\end{equation}
where the coefficients $\tilde{K}$ arise from integration over a different torsion one-cycle  $\tilde{\pi}_1$. 

We now impose the calibration condition. One can check that calibrated torsion two- and three-cycles in $\tilde{\bf T}^3 \times {\bf T}^3/\IZ_2 \times \IZ_2$ correspond to torsion one-cycles on $\tilde{\bf T}^3$ that are integral curves of $\chi$. For such one-cycles we have that $\int_{\pi_1} \alpha = \int_0^{2\pi r} \iota_\chi \alpha\, d\th$, for any one-form $\a$. Therefore 
\be
K_0 =  \int_0^{2\pi r} 	\iota_\chi  U_0 \, d\th = 2\pi r \phi_0 \, , \qquad K_\a^\pm =   \int_0^{2\pi r} 	\iota_\chi  U_\a^\pm \, d\th  \, .
\ee

From these expressions, one may compute each of the terms in the torsion linking number. Indeed, one first notices that
\begin{equation}
	\iota_\chi S_\a =   \lambda_\chi \phi_\a \,, \qquad \iota_\chi T_\a = ( \lambda_\chi^2 + \sigma_\a^2 -  \mu_\a^2) \, \phi_\a \,,
\end{equation}
which imply
\begin{equation}
	\iota_\chi U^\pm_\a = \pm \left\{\lambda^\pm_\a\bigg[ \lambda_\chi^2+\sigma_\a^2-\mu_\a^2\bigg]+\sigma_\a^2  \lambda_\chi\right\}\frac{c_\a^{\pm} }{\sqrt{ \lambda_\chi^2 + 4 \sigma_\a^2}} \phi_\a\,.
\end{equation}
As a result, the  massive eigenmodes with $\mu_\a \neq 0$ have a vanishing coefficient, since
\begin{equation}
	K^\pm_\a \propto \int_0^{2 \pi r}  \phi_\a \, d \th = 0  \,,
\end{equation}
where we have used \eqref{eqphiexp}. It remains to check the contribution of the modes with $\mu_\a = 0$ to \eqref{LtT3}. Recall that those modes with $\alpha\neq 0$ come in pairs, and one can check that they satisfy the following relation:
\begin{equation}
	\frac{1}{\lambda_\a^+} K^+_\a \tilde{K}^+_\a + \frac{1}{\lambda_\a^-} K^-_\a \tilde{K}^-_\a = \frac{\epsilon_\a}{ \lambda_\chi^2 + 4\sigma_\a^2} \int_0^{2 \pi r}  \phi_\a \, d\th \int_0^{2 \pi r}  \phi_\a \, d\th \,,
\end{equation}
where we have defined
\begin{equation}\label{eqdelta}
	\epsilon_\a \equiv \frac{(c_\a^{+})^2}{\lambda_\a^+} \left[\lambda_\a^+( \lambda_\chi^2+\sigma_\a^2) + \sigma_\a^2  \lambda_\chi\right]^2 + \frac{(c_\a^{-})^2}{\lambda^-_\a} \left[\lambda_\a^-( \lambda_\chi^2+\sigma_\a^2) + \sigma_\a^2  \lambda_\chi\right]^2 = 0 \,.
\end{equation}
That is, those massive eigenmodes with $\mu_\a =0$ have non-trivial coefficients $K_\a^\pm$, but for those contributing to the  bracket in \eqref{LtT3} there is a non-trivial cancellation by pairs, such that the sum cancels term by term. The surviving term in \eqref{LtT3} is the smeared linking number
\begin{equation}
	L^{\rm sm}(\pi_1,\tilde{\pi}_1) \equiv \frac{1}{ \lambda_\chi} K_0 \tilde{K}_0  =  \frac{4\pi^2r^2}{V_3  \lambda_\chi}\,.
 	\label{Lsmexc}
\end{equation}
In a twisted three-torus with metric $ds^2_{\tilde{\bf T}^3} = (2\pi)^2 \left[(R_i\eta^i)^2 + (R_{j+3}\eta^{j+3})^2 + (R_{k+3}\eta^{k+3})^2 \right]$ and twist $d\eta^i = - N \eta^{j+3} \wedge \eta^{k+3}$ one obtains 
\be
U_0 =\frac{2\pi R_i}{\sqrt{V_3}}\,\eta^i ,\qquad\lambda_\chi = \frac{NR_i}{2\pi R_{j+3}R_{k+3}}\, , \qquad r^2 = R_i^2\, , \qquad V_3 = 8\pi^3 R_iR_{j+3}R_{k+3} \, .\label{eq: U0T3}
\ee
Therefore applying \eqref{Lsmexc} one recovers the result $L^{\rm sm}(\pi_1,\tilde{\pi}_1) = 1/N$, as expected.

\subsubsection*{Extension to $\tilde{\bf T}^6/\IZ_2 \times \IZ_2$}

Let us now see how the above computation extends to the SU(3)-structure manifold $\tilde{\bf T}^6/\IZ_2 \times \IZ_2$. We first consider the covering space  $\tilde{\bf T}^6 = \tilde{\bf T}^3 \times {\bf T}^3$ with metric \eqref{SU3exc}, where $\tilde{\bf T}^3$ is parametrised by the coordinates $\{x^i, x^{j+3}, x^{k+3}\}$ and $ {\bf T}^3$ by $\{x^{i+3}, x^j, x^k\}$,  with $i \neq j\neq k \neq i$. Given the factorisation of the metric,  any eigenform of the Laplacian will be a wedge product of one in $\tilde{\bf T}^3$ and one in $ {\bf T}^3$. We are in particular interested in those eigenforms in which the bump delta-forms $\delta(\Pi_3^{\rm tor})$ and $\delta(\Pi_2^{\rm tor})$ are decomposed. It is easy to see that these fall in the subset
\bea
\left[ \star U_\a^\pm\right] & \wedge & \left(e^{2 \pi i n_{i+3} \, x^{i+3}}dx^{i+3}\right), \qquad n_{i+3} \in \IZ\, , \\
\left[ \star U_\a^\pm\right]  & \wedge & \left(e^{2 \pi (n_j  x^j+ n_k x^k)} dx^j \wedge dx^k \right), \qquad n_j, n_k \in \IZ\, ,
\eea
for  $\delta(\Pi_3^{\rm tor})$ and $\delta(\Pi_2^{\rm tor})$, respectively, where as above $\star$ stands for the Hodge star operator in $\tilde{\bf T}^3$. As a consequence,  the expansion of the smeared deltas $\delta^{\rm{sm}}(\Pi_3^{\rm tor})$ and $\delta^{\rm{sm}}(\Pi_2^{\rm tor})$ are given by $\star U_0\wedge dx^{i+3} $ and $\star U_0\wedge dx^{j}\wedge dx^k$, accordingly. That is, using the metric \eqref{SU3exc} one obtains
\be
  \delta^{\rm{sm}}(\Pi_3^{\rm tor}) = \eta^{4} \wedge \eta^{5} \wedge \eta^{6}  % dx^{i+3}\wedge dx^{j+3}\wedge dx^{k+3}, 
  \qquad
   \delta^{\rm{sm}}(\Pi_2^{\rm tor}) = -\eta^{j}\wedge \eta^{k}\wedge \eta^{j+3}\wedge \eta^{k+3}. \label{deltasmT6}
\ee
With regard to the complete expansion, it is easy to see that the wedge of one of these forms and its antiderivative will give a non-vanishing contribution only if $n_j = n_k = n_{i+3} =0$, that is if we select harmonic forms in ${\bf T}^3$. As a result, the computation of the linking number for calibrated cycles works precisely as outlined for $\tilde{\bf T}^3$, with the same vanishing coefficients and the same cancellations, and we end up again with the smeared torsion linking number \eqref{Lsmexc}. 

Let us now implement the $\IZ_2 \times \IZ_2$ orbifold projection, where each $\IZ_2$ generator $\th_1$ and $\th_2$ acts by flipping two coordinates on $\tilde{\bf T}^3$ and other two on $\tilde{\bf T}^3$, as follows from footnote \ref{ft:orbifold}. Since this is a product of two involutions, each acting on one submanifold, we can split the above exact eigenforms into even and odd under such involutions, and take (odd, odd) or (even, even) products, such that the result is invariant under the orbifold generators. While one could perform such an analysis explicitly, given our discussion above it is sufficient to show the action of these orbifold generators on the two-forms $\star U_\a^\pm$ only depends on the value of $\sigma_\a$ and $\mu_\a$, since then the orbifold projection will commute with relations that lead to the cancellations \eqref{eqdelta}, and they will also happen for orbifold-invariant massive modes. One can  show the assumption by using that $\th_1$ and $\th_2$ act as isometries when restricted to $\tilde{\bf T}^3$, as then they commute with $\Delta$, and ${\cal L}_\chi$. It then follows that they have a well-defined action on the basis of scalar wavefunctions $\{\phi_\a\}$, and act on the above set of co-exact one forms as
\be
\th_\a : S_\a \mapsto \nu_\a^{\sigma_\a, \mu_\a} S_\a   \, , \quad \th_\a : T_\a \mapsto \nu_\a^{\sigma_\a, \mu_\a} T_\a\  \implies \ \th_\a : U_\a^\pm  \mapsto \nu_\a^{\sigma_\a, \mu_\a}  U_\a^\pm\, .
\ee
That is, the orbifold group action on the massive modes of interest only depends on the value of $\sigma_\a$ and $\mu_\a$, as assumed. Finally, by construction, the orbifold projection leaves invariant the eigenmodes of $\tilde{\bf T}^3 \times {\bf T}^3$ that contribute to the smeared linking number.

%%%%%%%%%%%%%%%%%%%
%%%%%%%%%%%%%%%%%%%

\section{More general $\CN=2$ compactifications}
\label{s:general}

The extension of the setup in \cite{Gurrieri:2002wz} to more general type II string compactifications leading to 4d $\CN=2$ gauged supergravities has been performed in \cite{DAuria:2004kwe,Grana:2005ny,Kashani-Poor:2006ofe,Grana:2006hr}. In the following we focus on the framework developed in \cite{Grana:2005ny,Kashani-Poor:2006ofe}, which applies to SU(3)-structure manifolds. The results of \cite{Grana:2005ny,Kashani-Poor:2006ofe} imply that, on an SU(3)-structure manifold  $X_6$, one should be able to select a set of smooth $p$-forms:
\be
\{\om_a\} \in \Om^2(X_6)\, , \qquad \{\a_A, \b^B\} \in \Om^3(X_6)\, , \qquad \{\tilde{\om}^a\} \in \Om^4(X_6)\, , 
\ee
with $a=1, \dots, n_K$, $A,B = 1, \dots, n_{\rm c.s.}$, chosen such that 
\be
\int_{X_6} \om_a \wedge \tilde{\om}^b = \delta_a^b , \qquad \int_{X_6}  \a_A \wedge \b^B = \delta_A^B ,
\label{intrel}
\ee
and satisfying the relations 
\bes
\label{system}
\begin{align}
\label{coma}
d^\dag \om_a &= 0 , \\
  d\om_a &=  m_a{}^A \a_A + e_{aA} \b^A , \\
d\a_A &= e_{aA} \tilde{\om}^a , \\
 d\b^B &=  -m_a{}^B \tilde{\om}^a , \\
d\tilde{\om}^a & =  0 ,
\label{doma}
\end{align}
\ees
with $m_a{}^A, e_{aA} \in \IZ$ such that $m_a{}^A e_{bA} = m_b{}^A e_{aA}$. Consistency of the dimensional reduction implies %$\om_a \wedge \a_K = 0 = \om_a \wedge \b^K$, $\forall a,K$ and 
that the set is closed under the Hodge star operator:
\be
\tilde{\om}^a =  g^{ab} * \om_b , \qquad * \a_A = H_A^B \a_B + G_{AB}\b^B, \qquad * \b^A = F^{AB} \a_B - H^A_B \b^B ,
\label{hodgen}
\ee
mimicking the relations between harmonic forms in Calabi--Yau manifolds. 

Given this set of $p$-forms, one expands the SU(3)-structure calibrations in terms of them: 
\bea
\label{JcandOmgen}
J_c  &= &B+ iJ = 4\pi^2 (b^a + i t^a) \, \om_a , \\ \nonumber
\Om  &=  & Z^A\a_A - \cF_B\b^B ,
\eea
with $\cF_A = \p_A \cF$ the derivatives of the complex structure prepotential $\cF$. The 4d kinetic terms of the corresponding fields are governed by the same expressions as in the Calabi--Yau case, in terms of K\"ahler potentials  $K_\rho = - \log \int_{X_6} i \bar{\Om} \wedge \Om$ and $K_J = - \log \frac{4}{3}\int_{X_6} - J \wedge J \wedge J$ that correspond to Hitchin functionals  \cite{Grana:2005ny}. Finally, one should also expand the 10d RR potentials in this set of $p$-forms. In the case of type IIA compactifications such an expansion reads
\be
 C_3 = 2\pi \ell_s^3  \left(A_1^a  \wedge \om_a + \tilde{C}_0^A \a_A + C_{0\, B} \b^B\right) ,
 \label{C3gen}
\ee
leading to a set of axions and gauge vectors in 4d. The dual degrees of freedom are obtained from the expansion of $C_5$.

In the framework of \cite{Grana:2005ny,Kashani-Poor:2006ofe} there is no geometric interpretation for the set $\{\om_a, \a_A, \b^B, \tilde{\om}^b\}$, nor a clear prescription on how to build them from the light eigenmodes of the Laplacian. Notice that a key property of these $p$-forms is that they define the quantisation features of the 4d EFT, either in terms of axion periodicities or $U(1)$ gauge transformations. As such, their definition should be connected to the presence of 4d EFT objects like strings, particles and instantons, which implement and detect global gauge transformations. We have already seen this connection in our discussion of section \ref{s:dimred}, in light of which one may propose to describe the set $\{\om_a, \a_A, \b^B, \tilde{\om}^b\}$ as smeared delta forms. 

Indeed, based on our previous discussion, it is natural to propose that the smooth $p$-forms $\{\om_a, \a_A, \b^B, \tilde{\om}^b\}$ correspond to smeared delta-forms $\delta^{\rm sm}_p(\Sigma_{6-p})$ of a set of strictly calibrated $(6-p)$-chains $\Sigma_{6-p} \subset X_6$, which encode the presence of BPS objects in the 4d EFT. More precisely, the closed four-forms $\tilde{\om}^b$ correspond to the smeared bump delta-forms $\delta^{\rm sm}_4(\Pi_2)$, where $[\Pi_2]$ belongs to the free part of $H_2(X_6, \IZ)$ if $\tilde{\om}^b$ is harmonic, and to ${\rm Tor} H_2(X_6, \IZ)$ if it is de Rham exact. Similarly, the subset of three-forms in $\{ \a_A, \b^B\}$ that are closed correspond to the smeared delta-forms $\delta^{\rm sm}_3(\Pi_3)$ of strictly calibrated three-cycles. The remaining set of smooth $p$-forms can be constructed by taking the anti-derivatives of the exact three- and four-forms and normalising them such that \eqref{intrel} is satisfied, which implies that the integers $m_a{}^A, e_{aA}$ encode the torsion linking numbers of $X_6$. Finally, as in our simple example above, one could also relate the non-closed two- and three-forms as the smeared version of delta-forms for calibrated four- and three-chains in $X_6$, whose boundary describes the torsional nature of some calibrated cycles. 

To make this picture more precise, let us consider the subcase $m_a{}^A =0$, which also resembles the setup considered in \cite{Camara:2011jg}. Then, the rank $r_{\bf e}$ of the matrix $e_{aA}$ should determine the number of harmonic two- and three forms of $X_6$ as $b_2(X_6) = n_K- r_{\bf e}$ and $b_3(X_6) = n_{\rm c.s.} - r_{\bf e}$. Clearly, the rank of $e_{aA}$ counts massive  eigenforms below the compactification scale, more precisely we should at least have $r_{\bf e}$ times a spectrum of the form \eqref{massivepforms}, as this is what we obtain from dimensionally reducing the RR sector of the theory. Indeed, let us consider the type IIA expansion \eqref{C3gen}, and for simplicity assume that in \eqref{hodgen} $H^A_B =0$, so that $G_{AB}F^{BC} = - \delta_A^C$. Then we find
\be
(2\pi)^2 \hat{F}^{AB} \left( dC_{0\,  A} - e_{aA} A_1^a\right) \left( dC_{0\, B} -e_{bB} A_1^b\right) + (2\pi)^2 g_{ab}\, dA_1^a \wedge dA_1^b\, , 
\label{disgaugen}
\ee
with $ \hat{F}^{AB} =   F^{AB} e^{2\phi_4} M_{\rm P}^2$, plus a mass term for $r_{\bf e}$ axions $\tilde{C}_0^A$.\footnote{These massive axions are more suitably described in terms of a 4d dual two-form $B_2$ involved in a gauging of the form \eqref{p=2}, see e.g. \cite{Marchesano:2014mla}.} The masses that one reads from such a mass term, the Lagrangian \eqref{disgaugen} and its dual reproduce the action of the Laplace operator on the set $\{\om_a, \a_A, \b^B, \tilde{\om}^b\}$ as expected \cite{Tomasiello:2005bp,Grana:2005ny,Kashani-Poor:2006ofe}. For instance, the action of the Laplacian on the closed forms $\b^A$ and $\tilde{\om}^a$ reads
\begin{eqnarray}
    \Delta \beta^A &= &F^{AB}e_{bB}g^{bc}e_{cC}\, \beta^C \, , \\
    \Delta \tilde{\om}^a &= & g^{ab} e_{bB}F^{BC}e_{cC} \, \tilde{\om}^c \, .
\end{eqnarray}
The diagonalisation of these mass matrices gives us the set of massless and light $p$-form eigenmodes. Such a spectrum is by assumption complete, or otherwise the expansions \eqref{JcandOmgen} and \eqref{C3gen} would be missing light modes of the EFT.     Knowledge of these mass matrices and of the kinetic terms $C^{AB}$ and $g_{ab}$ leads to $e_{aA}$, in a generalisation of the relation \eqref{Lphys}. As proposed in \cite{Camara:2011jg}, the matrix $e_{aA}$ is a sort of the inverse of the torsion linking numbers, and it encodes the torsion cohomology that is sensitive to the light EFT modes. This topological information is easier to extract if one performs a unimodular integral change of basis both in $\{\om_a, \tilde{\om}^b\}$ and in $\{\a_A, \b^B\}$ that take $e_{aA}$ to its Smith normal form
\be
e^{\rm Smith} =
\begin{pmatrix}
k_1 \\ & k_2 \\ & & \ddots \\& & & k_{r_{\bf e}} & \dots & 0\\ & & & 0 &\dots & 0 \\  & & & \vdots &\ddots & \vdots \\ & & & 0 &\dots & 0
\end{pmatrix}
\label{eSmith}
\ee
with $k_i, k_i/k_{i+1} \in \IZ, \forall i$. In this basis the computation of the smeared linking number gives $\int_{X_6} d^{-1} \tilde{\om}^i \wedge \beta^j = k_{i}^{-1} \delta^{ij}$, for $i,j = 1, \dots, r_{\bf e}$, suggesting that the torsion cohomology groups are
\be
{\rm Tor }\, H^3(X_6, \IZ) \simeq {\rm Tor }\, H^4(X_6, \IZ) \simeq \IZ_{k_1} \times \dots \times \IZ_{k_m}\, ,
\label{torsmith}
\ee
where $|k_i| > 1$ for $i \leq m$ and $|k_i| = 1$ for $m< i \leq r_{\bf e}$. Those entries of \eqref{eSmith} with value $\pm1$ should correspond to calibrated $p$-cycles that are trivial in homology, but that nevertheless are detected by the 4d EFT because they couple to massive modes below the compactification scale.

The proposal that the closed $p$-forms within $\{\om_a, \a_A, \b^B, \tilde{\om}^b\}$ correspond to smeared bump delta-forms of calibrated cycles can be further motivated by considering the set of BPS objects in the 4d EFT. For instance, let us again consider type IIA with $m_a{}^A =0$ and look at the closed three-forms $\b^A$, which in the basis \eqref{eSmith} may either be harmonic or exact in de Rham cohomology. Each of these forms are related to an axion $C_{0\, A}$, and from the BPS completeness hypothesis \cite{Polchinski:2003bq}, or the EFT string completeness hypothesis \cite{Lanza:2021udy} applied to $\CN=2$ gauged supergravities, one expects a BPS string under which such an axion is magnetically charged. Then, the results of \cite{Grana:2005ny,Kashani-Poor:2006ofe} imply that $K_\rho = - \log \int_{X_6} i \bar{\Om} \wedge \Om$  describes the metric of the hypermultiplet moduli space, at least at the classical level. Because the axion kinetic terms only depend on $K_\rho$ and this has the same expression as in the ungauged case, the tension of a BPS string should have the same general expression as in a Calabi--Yau. That is, we have that
\be
\frac{{\cal T}^A}{M_{\rm P}^2} = \frac{e^{\phi}}{{\rm Vol}_{X_6}} \left| \int_{X_6} \Om \wedge \beta^A \right| \, .
\ee
Now, in this context BPS means that the D4-brane internal worldvolume $\Pi_3^A$ is calibrated by $\Omega$, and dimensionally reducing its DBI action one obtains
\be
\frac{{\cal T}^A}{M_{\rm P}^2} =  \frac{e^{\phi}}{{\rm Vol}_{X_6}}  \left| \int_{\Pi_3^A}  \Om \right| =  \frac{e^{\phi}}{{\rm Vol}_{X_6}}  \left| \int_{X_6}  \Om \wedge \delta(\Pi_3^A) \right| ,
\ee
which implies that $\b^A$ must be the smeared version of $\delta(\Pi_3^A)$. Note that this is a standard result when $\b^A$ is not exact in de Rham cohomology, since then $[\b^A]$ and $[\Pi_3^A]$ are related by standard Poincar\'e duality. Similarly, in 4d $\CN=2$ EFTs, the mass of charged BPS particles in Planck units is specified by their central charge. A key result of \cite{Grana:2005ny,Kashani-Poor:2006ofe} is that the kinetic terms of vector multiplet sector is encoded in the K\"ahler potential $K_J = - \log \frac{4}{3}\int_{X_6} - J \wedge J \wedge J$ also for gauged supergravities obtained from compactifications on SU(3)-structure manifolds. From here it follows that the central charges of BPS particles charged under the vector multiplets are precisely the periods of $B+iJ$, that is
\be
Z^a = e^{K_J/2} \int_{X_6} e^{B+iJ} \wedge \tilde{\om}^a .
\ee
As in the Calabi--Yau case, such BPS particles should arise from wrapping D2-branes on two-cycles $\Pi_2^a \subset X_6$  calibrated by $J$. By dimensionally reducing their DBI action we obtain
\be
\frac{m_a^2}{M_{\rm P}^2} = e^{K_J} \left| \int_{X_6} (B+i J)\wedge \delta (\Pi_2^a) \right|^2  ,
\ee
which implies that $\tilde{\om}^a$ should be the smeared version of $\delta(\Pi_2^a)$. Again, this is independent of whether $\tilde{\om}^a$ is de Rham exact or not. When $\tilde{\om}^a$ is exact, it has to be that $\Pi_2^a$ is either a torsion or a trivial class of $H_2(X_6, \IZ)$. Finally, the completeness hypothesis implies that there is a BPS particle per element of the basis $\{\tilde{\om}^a\}$, which again is a standard result in the Calabi--Yau case.

%%%%%%%%%%%%%%%%%%%
%%%%%%%%%%%%%%%%%%%

\section{Torsion D-branes in $\CN=1$ vacua}
\label{s:N=1}

Having discussed the physical meaning of the smeared torsion linking number in 4d $\CN=2$ settings, it is natural to wonder how Conjecture \ref{conj:BPS} can be physically realised in 4d $\CN=1$ string vacua. In the $\CN=2$ case, the realisation was based on the existence of BPS AB particles and strings, a set of objects that will be essentially absent in 4d $\CN=1$ type II orientifold settings. Indeed, there are no BPS particles in $\CN=1$ vacua, and a 4d string that arises from a D$(p+1)$-brane wrapped on a $p$-cycle $\Pi_p \subset X_6$ can only be BPS if $\Pi_p$ is calibrated by a closed $p$-form.\footnote{The precise statement is that in 4d $\CN=1$ Minkowski vacua the calibration for 4d strings is $d_H$-closed  \cite{Martucci:2005ht,Koerber:2005qi,Koerber:2010bx}, an statement that also holds for the $\CN=0$ Minkowski vacua analysed in \cite{Lust:2008zd}. In practice this implies that, even in compactifications with $H$-flux,  torsion cycles cannot be calibrated.} As such, torsion $p$-cycles cannot yield 4d BPS strings. The realisation of the smeared linking number must therefore be more subtle in this case. 

As already mentioned in section \ref{s:dimred}, one possibility is to invoke the extension of Conjecture \ref{conj:BPS} formulated around \eqref{deldif}, and look for torsion $p$-cycles that are not calibrated by themselves, but nevertheless its homology class can be seen as a linear combination of calibrated $p$-cycle classes. This would in principle allow us to describe AB strings and particles and their associated smeared $p$-forms in 4d $\CN=1$ orientifold vacua, connecting our previous discussion with the setup of \cite{Camara:2011jg}.%, although the description of AB particles in that setting would still remain unclear.

The most natural realisation of the smeared torsion linking form seems instead to involve the St\"uckelberg-like terms that involve 4d BPS objects in $\CN=1$ vacua, and that correspond to \eqref{p=2} and \eqref{p=3}. These two couplings represent 4d membranes ending on strings and space-time filling branes ending on membranes, and in the context of 4d $\CN=1$ supersymmetry they can be described in the language of three-form multiplets  \cite{Lanza:2019xxg}. Notice that in a type II compactification on a six-dimensional manifold $X_6$, a D$p$-brane that looks like a 4d particle and a D$(p+2)$-brane that looks like a 4d membrane can wrap the same $p$-cycle $\Pi_p \in X_6$, and the same holds for a D$(p+3)$-brane and an Euclidean  D$(p-1)$-brane instanton. So essentially we are trading the role of 4d particles and instantons for that of 4d membranes and space-time filling branes, in order to probe a similar set of torsion $p$-cycles with BPS objects. In practice, this implies that the topological information that was captured by \eqref{stuck} and \eqref{stuckdual} in the $\CN=2$ case, now is encoded in the couplings \eqref{p=2} and \eqref{p=3}.

To see how this works in practice, let us again focus on type IIA string theory on a compact SU(3)-structure manifold $X_6$, but now with an orientifold projection that introduces O6-planes. Assuming a 4d Minkowski vacuum leads to the following metric Ansatz
\be
ds^2 = e^{2A} ds^2_{\IR^{1,3}} + \ell_s^2 ds^2_{X_6} ,
\ee
with $A$ a warp factor that depends on the coordinates of $X_6$, whose SU(3)-structure metric satisfies the following equations
\be
d(3A-\phi) = H + idJ = 0 , \qquad d(e^{2A-\phi} {\re \Om}) =0 , \qquad \ell_s d( e^{4A-\phi} {\im \Om}) = - e^{4A} * F_2 ,
\label{minkIIA}
\ee
and $F_0 = F_4 =F_6 =0$, with $F_{2p}$ the gauge-invariant RR field strength. In this setup 4d strings made up of D4-branes wrapping three-cycles are calibrated by $\pm e^{2A-\phi}{\re \Om}$ which, as advanced, is a closed three-form. Therefore, there are no BPS strings of this sort that correspond to torsion homology classes. The same can be said for membranes, which are calibrated by $e^{3A-\phi}e^{B+iJ}$. 

The last equation in \eqref{minkIIA}, however, features a non-closed three-form that calibrates space-time filling D6-branes. As such, it can detect calibrated torsion three-cycles. That such three-cycles exist in certain SU(3)-structure vacua can be deduced from the results of \cite{Tomasiello:2005bp}, which imply that a D$p$-brane that is point-like in a Calabi--Yau manifold with $H$-flux is mapped by mirror symmetry to a D$(p+3)$-brane wrapping a torsion three-cycle $\Pi_3^{\rm tor}$. In our type IIA orientifold context, $\Pi_3^{\rm tor}$ will host a 4d BPS object if it is wrapped either by a D6-brane or by an Euclidean D2-brane. In practice, the DBI action of these objects is easier to analyse if, following \cite{Tomasiello:2007zq}, one trades the last equation in \eqref{minkIIA} by an equivalent one not involving the Hodge star operator. In the case at hand we find \cite{Marchesano:2014iea}
\be
\ell_s d( e^{-\phi} {\im \Om}) = - J \wedge F_2 ,
\label{IIAJF}
\ee
which already hints that part of the torsion data of $X_6$ is encoded in the RR flux $F_2$. 

In this context, it is illustrative to consider a simple example, like the twisted six-torus geometries analysed in \cite{Marchesano:2006ns}. These correspond to an SU(3)-structure of the form \eqref{Jex} and \eqref{Omex}, with the simplification $V= V_i =1$, $\forall i$, and an orientifold action of the form ${\cal R}: (J,\Om) \mapsto {-}(J,\bar{\Om})$. The $p$-chains $\{\Pi_2^i, \Pi_3^{\rm tor}, \Sigma_3 , \Sigma_4^i \}$ play the same role in terms of torsion homology information as in section \ref{s:simple}, but from an EFT viewpoint they should be associated to either 4d membranes or space-time filling branes. For concreteness, let us consider a $\tilde{\bf T}^6$ with twisting $M_1 = -M_2 = N \in \mathbb{N}$ and $M_3=0$. This choice fixes the complexified K\"ahler moduli as $b^1+it^1 = b^2+it^2$ in the vacuum and leads to a background RR flux of the form
\be
F_2 = \ell_s K \left(\eta^1 \wedge \eta^4 -\eta^2 \wedge \eta^5\right) , \qquad K \in \mathbb{N}\, ,
\label{F2orism}
\ee
that solves \eqref{IIAJF} in the constant dilaton approximation, by setting $K t^1 = K t^2 = N e^{-\phi}R_1R_2R_3$. This sort of RR flux background is the one that appears in the type IIA orientifold flux literature, see e.g. \cite{Villadoro:2005cu,Camara:2005dc}, but it is important to realise that the expression is a consequence of the constant-dilaton/smeared approximation. Indeed, what occurs in this background is that there are eight O6-planes wrapped on $[\Pi_3^{\rm tor}]$. In a plain toroidal compactification one could cancel this charge by placing 32 D6-branes in the same three-cycle class of the covering space. In the twisted torus geometry, because $[\Pi_3^{\rm tor}]$ is $\IZ_N$-torsion, one only needs to place $32 - kN$ of such D6-branes, for some $k \in \IZ$, in order  to cancel the RR tadpole. The lack of D6-branes leads to a RR flux background that satisfies
\be
dF_2 \simeq -\ell_s k N \delta_3 (\Pi_3^{\rm tor}) ,
\label{F2oriunsm}
\ee
where for simplicity we have assumed all O6-planes and D6-branes on the same representative (otherwise one is led to more involved delta-source equations, like the ones solved in \cite{Casas:2022mnz}). Upon implementing the smearing approximation one obtains
\be
F_2 = - \ell_s k N\, d^{-1}\delta_3^{\rm sm} (\Pi_3^{\rm tor}) \, ,
\ee
which reproduces \eqref{F2orism} for $k =2K$, up to a harmonic form. The actual RR flux is, however, the one that solves \eqref{F2oriunsm}, since it is the only one that can satisfy Dirac's quantisation condition, upon the appropriate choice of harmonic piece \cite{Marchesano:2014iea}. 

The couplings \eqref{p=2} and \eqref{p=3} are obtained upon dimensionally reducing the RR potentials
\bea
 C_5 & = &  \ell_s^{5} 2\pi \left[ B_{2\, 0} \wedge \b^0 + B_2^i \wedge \a_i  + C_3^i \wedge \om_i\right]\, , \\
 C_7 & = & \ell_s^{7} 2\pi \left[ D_{3\, i} \wedge \tilde{\om}^i + A_{4}^0 \wedge \a_0  + A_{4\, i}\wedge \b^i \right]\, ,
\eea
where we have taken into account the orientifold action, and expanded into $p$-forms that couple with unit charge to the calibrated $p$-chains $\{\Pi_2^i, \Pi_3^{\rm tor}, \Sigma_3 , \Sigma_4^i \}$.\footnote{For simplicity we are using the notation \eqref{alphabetas}, which results in an unusual convention in orientifold compactifications. The more standard one is obtained by interchanging the basis elements as $\a_i \leftrightarrow - \b^i$, cf. \cite[eq.(2.6)]{Camara:2005dc}.} One finds 
\be
(2\pi)^2 \left[ \hat{g}_{ii} (dC_3^i)^2 + \tilde{F}^{00}  \left(d B_{2\, 0} + M_i C_3^i \right)^2 + e^{-4\phi_4}M_P^{-4} \tilde{F}_{ii}  (dB_2^i)^2 \right]\, ,
\label{p=2ori}
\ee
and
\be
(2\pi)^2 \left[ \hat{g}^{ii} \left( d D_{3\, i} - M_i A_4^0 \right)^2 \right]  e^{-8\phi_4}M_P^{-8} \, ,
\label{p=3ori}
\ee
where we have assumed generic twists $M_i$ and have defined
\be
\hat{g}_{ii}=g_{ii} 
e^{-4\phi_4}M_P^{-4},\qquad
\tilde{F}^{AA} = F^{AA} 
e^{-2\phi_4} M_P^{-2}, 
\ee
and $\hat{g}^{ii}=1/\hat{g}_{ii}$, $\tilde{F}_{AA}=1/\tilde{F}^{AA}$. Note that $\hat{F}^{AB}$ defined below \eqref{disgaugen} satisfies $\hat{F}^{AB} = \tilde{F}^{AB} e^{4\phi_4}M_P^{4}$, and so \eqref{p=2ori} contains the same kind of information as \eqref{StuckN=2gen}. As a result, the computation of the smeared linking number works exactly as in section \ref{ss:EFTdesc}. The main difference is the expression of the smeared linking number in terms of 4d EFT quantities, which now involves the physical charges of membranes and strings ending on each other. More precisely in this $\CN=1$ setup, one finds that the relation \eqref{Lphys} is substituted by
\be
  \frac{m_{\rm st}}{N} =  \sqrt{\frac{\hat{g}^{\a\a}}{\tilde{F}_{00}}} = e^{\phi_4} M_{\rm P} \sqrt{\frac{{g}^{\a\a}}{{F}_{00}}}\, ,
\ee
where $N=M = {\rm g.c.d.} (M_1, M_2, M_3)$ and $\hat{g}^{\a\a} = e^{4\phi_4}M_P^{4} {g}^{\a\a} = e^{4\phi_4}M_P^{4} \sum_i \frac{M_i^{2}}{M^2} g^{ii} $. Here $\hat{g}^{\a\a}$ represents the squared physical charge ${\cal Q}^2$ of a BPS membrane ending on a BPS string and $\tilde{F}_{00}$ the squared charge of such a string, as defined in \cite{Lanza:2020qmt}, see also \eqref{physQ}.

%%%%%%%%%%%%%%%%%%%
%%%%%%%%%%%%%%%%%%%

\section{Beyond the BPS case}
\label{s:nonBPS}

Conjecture \ref{conj:BPS} proposes a method to compute the linking number between two calibrated torsion cycles from smeared/EFT data. However, as already mentioned, in Calabi--Yau manifolds torsion $p$-cycles cannot be calibrated, or equivalently D-branes wrapped on them are not BPS objects of the EFT. The extension of the conjecture around \eqref{deldif} allows us to implement the same method whenever the torsion class $\Pi_p^{\rm tor}$ of interest is a linear combination of $p$-cycle homology classes with calibrated representatives. This more general setup could in principle occur in Calabi--Yau compactifications, and then the extended conjecture would imply that one can compute torsion in cohomology via smeared data, provided there exist massive eigenforms of the Laplacian below the compactification scale that couple to torsion $p$-cycles. Including such a set of light fields in the 4d EFT would presumably take us to a structure of the form \eqref{intrel} and \eqref{system}, in which giving a non-vanishing vev to a massive, light field deforms an SU(3)-holonomy metric to an SU(3)-structure one.  

Nevertheless, in general, one would expect that a torsion class in homology does not contain any calibrated representative, and neither can it be understood as a linear combination of homology classes with them. In that case, our discussion of section \ref{s:dimred} suggests that there should be non-trivial corrections  associated to this sector. More precisely, one would expect that in the EFT description the bump-delta form $\delta_{6-p} (\Pi_p^{\rm tor})$ can still be replaced by its smeared version, but only  up to a multiplying constant that could be interpreted as  wavefunction renormalisation. That is, one does not simply project the delta into its lowest eigenmode component, but also has to multiply the result by some constant (or a field-dependent function) in order to correctly reproduce the 4d physics. Whenever this happens the smeared linking number and the exact linking number do not coincide, and one should rescale the smeared delta to make it so. 

It is hard to have an idea of the magnitude of this rescaling without an  example at hand where the computation can be carried out explicitly. The best we can do is to give an estimate for the error in the smeared linking number, as follows. In supersymmetric theories, the EFT kinetic terms obtained from truncating  massive modes at zero vev are exact up to corrections of ${\cal O}(\Lambda_{\rm EFT}/\Lambda_{\rm UV})$, see e.g. \cite{Buchmuller:2014vda}. In our case $\Lambda_{\rm EFT}$ corresponds to the mass $m_{\rm st}$ of the massive modes that we keep in our EFT, and $\Lambda_{\rm UV}$ to the compactification scale $m_{\rm KK}$. Notice that this is the suppression that we found in \eqref{shiftKK} when moving away from the minimal tension representative. 

If the torsion cycle is not calibrated, it means that the EFT data is not computing its minimal volume properly. In some cases, like in Calabi--Yau vacua the `expected' volume ${\cal V}_p^{\rm BPS}$, namely the integral of the appropriate calibration over $\Pi_p^{\rm tor}$,  vanishes. In general, the volume of a non-calibrated cycle should be larger than the integral of its would-be calibration over the given homology class. One can think of the mismatch between volumes as how much one needs to deform a would-be calibrated cycle to match the actual one. Finally, one usually converts differences of internal volumes to differences of field vevs via the physical 4d charge ${\cal Q}$ of the corresponding EFT object  \cite{Lanza:2021udy}, which here we define as 
\be
{\cal Q}^2(\Pi_p) = \sum_{\lambda_i \ll \ell_s m_{\rm KK}} c_i^2 ,
\label{physQ}
\ee
with $c_i$ the coefficients of the smeared delta-form of $\Pi_p$, as in \eqref{deltasm}.  

With these considerations in mind, let us consider the smeared linking number between a calibrated cycle $\Pi_{6-p-1}^{\rm tor}$ and a non-calibrated one $\Pi_p^{\rm tor}$. One may propose the following upper bound 
\be
\left| L- L^{\rm sm} \right| < \frac{{\cal V}
_p - {\cal V}_{p}^{\rm BPS}}{{\cal Q}} \frac{m_{\rm st}}{m_{\rm KK}} ,
\ee
where ${\cal V}$ is the actual volume of $\Pi_p^{\rm tor}$, in string units. This upper bound estimates the error when computing the smeared linking number, with respect to the actual one $L$. If the bound is small, it still makes sense to compute \eqref{Lsm}, because it gives a good estimate of the actual linking number. That is, one may still use EFT data to characterise torsion in cohomology.

%%%%%%%%%%%%%%%%%%%
%%%%%%%%%%%%%%%%%%%

\section{Conclusions}
\label{s:conclu}

In this work we have proposed a method to detect topological invariants of torsion cohomology groups via smooth $p$-forms. The proposal is based on what ${\rm Tor} H^p(X_n, \IZ)$ means when performing dimensional reduction of type II string theory on $X_n$ and obtaining a lower-dimensional EFT with a massive sector, and it can be summarised in two main points: 

\begin{itemize}

\item[{\it i)}] If a D-brane wrapped on a torsion cycle $\Pi_p^{\rm tor}$ has a non-trivial backreaction at EFT wavelengths, it is because there are light massive eigenmodes of the Laplacian sourced by it. In geometric terms, this means that $\Pi_p$ has a non-trivial smeared delta form $\delta^{\rm sm}_{n-p} (\Pi_p)$, which is a necessary requirement to apply our approach.

\item[{\it ii)}] Whenever $\Pi_p^{\rm tor}$ is calibrated, a D-brane wrapped on it is a BPS object of the theory whose smeared backreaction is protected from  dimensional reduction corrections. As a result one can compute the torsion linking numbers of $\Pi_p^{\rm tor}$ using its smeared delta form. 
\end{itemize}

This second statement, which is the content of Conjecture \ref{conj:BPS}, provides a method to detect the $\IZ_N$ factors in $H^p(X_n, \IZ)$. The method has a wider application if one assumes the extension of the conjecture made around \eqref{deldif}, and it would be really interesting to see if it can be applied to manifolds with special holonomy metrics. 

The use of smooth $p$-forms to compute torsion in cohomology may seem quite surprising because such $\IZ_N$ factors are projected out in de Rham cohomology groups. One should however keep in mind that in our approach we are starting with a set of objects that contain the information of singular homology groups, namely the bump delta forms $\delta_{n-p} (\Pi_p)$, and replacing them by a countable set of smooth forms $\delta_{n-p}^{\rm sm} (\Pi_p)$ that should remember part of the torsion data. As a possible analogy, one may consider a finite good cover of $X_n$ and its nerve $N$, which is a triangulation of $X_n$ \cite{Bott1982DifferentialFI}. We then consider the delta forms $\delta_{n-p} (\sigma_{p, \a})$ of the $p$-simplexes $\sigma_{p, \a}$ of $N$, with a small smearing (such that $1/\lambda_{\rm max}$ is below the spacings in $N$). This produces a lattice of smooth $p$-forms from which one can compute $H^p(X_n, \IZ)$ via singular cohomology. Our proposal can be thought of as a limit of this construction, in the sense that we perform a much more dramatic smearing, namely at wavelenghts above ${\rm Vol}(X_n)^{1/n}$.  This more drastic coarse-graining is allowed geometrically because, if the assumptions behind Conjecture \ref{conj:BPS} are true, then there should be a $G$-structure manifold that one can construct by fibering $X_n$ over flat space, which is the EFT solution of a D-brane wrapped on $\Pi_p$. One could then use this non-compact, higher-dimensional manifold to compute topological information of $X_n$ via smeared data. 

It is also instructive to compare our approach with some of the discussion in the string theory literature, like the one carried in \cite{Tomasiello:2005bp} based on the classification theorems of Wall \cite{WALL1966} and \v Zubr \cite{Zhubr1978ClassificationOS}. As pointed out in \cite{Tomasiello:2005bp} these theorems classify six-manifolds up to diffeomorphisms and the classification data match the content of the massless sector of the compactification, discarding exact and co-exact $p$-forms. Our results are not in tension with this classification, because we need to endow $X_n$ with a $G$-structure metric in order to extract the torsion cohomology data. This choice of metric also specifies the set of calibrated cycles and the light spectrum of the Laplacian, so it is crucial in order to select those exact and co-exact forms that contain the torsion information. In this light, it would be interesting to see if the presence of torsion in cohomology restricts the choice of $G$-structure metrics on a manifold, or if one can always choose a $G$-structure metric where all the smeared delta forms of calibrated torsion cycles vanish. 

An important part of our analysis is based on constructing explicit examples of SU(3)-structure manifolds with calibrated torsion cycles. This allowed us to perform a direct comparison of the torsion linking number and its smeared version, where we observed a remarkable cancellation between terms in the eigenmode expansion of the delta form, that is reminiscent of the computations of topological indices. It would be very interesting to understand the meaning of this feature and if it is also realised in more involved setups, providing further evidence of Conjecture \ref{conj:BPS}. Our explicit constructions also provided concrete EFT descriptions of BPS configurations of branes ending on branes, like the 4d Aharanov-Bohm strings stretching between a domain wall and a monopole in $\CN=2$ gauged supergravities. Moreover, the properties of such objects resulted in a physical interpretation of the Hitchin flow equations, which could be useful to further understand the properties of these subtle objects. It is likely that this connection sheds light into the physics of $\CN=2$ gauged supergravities, like for instance when applied to the black hole supergravity solutions recently revisited in \cite{Angius:2023xtu}, and which share many properties with AB strings and particles. 

As a direct application of our proposal, we have revisited the dimensional reduction framework developed in \cite{Gurrieri:2002wz,DAuria:2004kwe,Grana:2005ny,Kashani-Poor:2006ofe} to furnish it with one of its main missing elements. That is, a geometric prescription to define the basis of $p$-forms in which the RR potentials and the calibration forms must be expanded. We have verified that our definition fits perfectly with the physical properties that these forms should have, and which define the periodicity properties of massive axions and gauge bosons of the 4d $\CN=2$ EFT. Such periodicities are crucial to define the global gauge transformations for massive $p$-form in more general setups. This applies in particular to  4d $\CN=1$ compactifications, where the St\"uckelberg-like couplings related to our method involve the gauging of three- and four-forms in 4d. 

To sum up, our findings seem to point out that torsion in cohomology could lead to specific, measurable physics in the massive sector of 4d EFTs obtained from string theory. It could be that exploiting this new link between geometry and physics could give us a new, more approachable understanding of the subtle objects that are torsion $p$-cycles.

%%%%%%%%%%%%%%%%%%%
%%%%%%%%%%%%%%%%%%%

\bigskip

\centerline{\bf  Acknowledgments}

\vspace*{.5cm}

We would like to thank I\~naki Garc\'ia-Etxebarria, Luca Martucci, Luca Melotti, Miguel Montero, David Prieto and \'Angel Uranga for discussions.  This work is supported through the grants CEX2020-001007-S and PID2021-123017NB-I00, funded by MCIN/AEI/10.13039/501100011033 and by ERDF A way of making Europe. G.F.C. is supported by the grant PRE2021-097279 funded by MCIN/AEI/ 10.13039/501100011033 and by ESF+. M.Z. is supported by the fellowship LCF/BQ/DI20/11780035 from ``La Caixa" Foundation (ID 100010434).

%%%%%%%%%%%%%%%%%%%%%%%%%%%%%%%%%%%%%%%%%%%%%%%%%%%%%%%%%%%%%%%%%%%%%%%%%%%%%%%%%%%%%%%%%%%%%%%%%%%%%%%%%%%%%%%%%%%%%%%%%%%%%%%%%%%%%%%%%%%%%%%%%%%%%%%%%%%%%%%%%%%%%%%%%%%%%%%%%%%%%%%%%%%%%%%%%%%%%%%%%%%%%%%%%%%%%%%%%%%%%%%%%%%%%%%%%%%%%%%%%%%%%%%%%%%%%%%%%%%%%%%%

\appendix

\section{NS5-branes and domain wall solutions}
\label{ap:NS5DW}

In  \cite{Gurrieri:2002wz} a 4d domain-wall solution based on a backreacted N5-brane is considered, in order to arrive at a background of the form \eqref{SU3ex} via mirror symmetry. In this appendix we review the NS5-brane setup, and describe a set of type IIB D-branes that are mutually BPS with the NS5-brane. Such D-branes have a well-defined interpretation from the 4d domain-wall viewpoint, and are in one-to-one correspondence with the type IIA D-brane objects discussed in section \ref{s:simple}. 

Following  \cite{Gurrieri:2002wz}, let us consider a toroidal compactification of type IIB string theory to 4d. Let us call the three complex coordinates of ${\bf T}^6 = ({\bf T}^2)_1 \times ({\bf T}^2)_2 \times ({\bf T}^2)_3$ as $dz^j = dx^j + i dx^{j+3}$, $j=1,2,3$. Then one places $M$ NS5-branes along the three-cycle $(1,0)_1(1,0)_2(0,1)_3$ (that is, the coordinates $\{x^1, x^2, x^6\}$) and spanning two spatial dimensions in $\IR^{1,3}$. One then backreacts such NS5-branes and smears the solution down to 4d. In this approximation the harmonic function $V$ that describes an NS5-brane backreaction becomes a linear function of its transverse coordinate in 4d, therefore $V = 1- \zeta \xi$ with $\zeta$ a real constant.\footnote{See footnote \ref{ft:dw} for an explanation of different choice of $V$ compared to \cite{Gurrieri:2002wz}.} One obtains the background
\bea
\label{Hex}
ds^2 &= &ds^2_{\IR^{1,2}}+ \ell_s^2 V(d\xi)^2 + \ell_s^2 ds^2_{{\bf T}^6}, \\
ds^2_{{\bf T}^6} & = & (2\pi)^2 \left[ (r_1 dx^1)^2+(r_2 dx^2)^2+V(r_3 dx^3)^2+V( r_4 dx^4)^2+V(r_5 dx^5)^2+(r_6 dx^6)^2 \right] , \\
H & = & - M dx^4 \wedge dx^5 \wedge dx^3, \\
e^{2\phi} & = & e^{2\phi_0} V .
\eea
with $M \in \mathbb{N}$. Upon three T-dualities along the coordinates $\{x^1, x^2, x^3\}$ one is led to a type IIA background with constant dilaton, no $H$-flux, and an internal metric of the form \eqref{SU3exb}, more precisely with $M_3=M$ and $R_j = 1/r_j$ for $j=1,2,3$. 

This smeared solution is interpreted as the long-wavelength description of the $M$ NS5-branes'  backreaction, more precisely as the 4d domain-wall solution that is perceived at wavelengths much larger than the compactification radii. As such, a D-brane that is BPS in the microscopic background should also be so in its long-wavelength approximation. In practice, this means that if we consider D-branes that are mutually BPS with the NS5-branes sourcing the solution, they should correspond to BPS objects in the 4d domain-wall background. To match out discussion with that of section \ref{s:simple}, we will consider type IIB D-branes whose embedding survives the $\IZ_2 \times \IZ_2$ orbifold projection with generators $\theta_1: (x^1, x^2, x^3, x^4, x^5, x^6) \mapsto  (x^1, -x^2, -x^3, x^4, -x^5, -x^6)$ and $\theta_2: (x^1, x^2, x^3, x^4, x^5, x^6) \mapsto  (-x^1, -x^2, -x^3, -x^4, -x^5, x^6)$.

For instance, D1 and D5-branes are mutually supersymmetric with respect to an NS5-brane if the system has $2+4k$ ND directions.\footnote{In an abuse of language, we are borrowing the nomenclature used for configurations of pairs of D-branes.} We may then consider

\begin{itemize}

\item[-] A D5-brane wrapping  $(T^2)_i \times (T^2)_j$ and extended along $\xi$  $\Longrightarrow$ 6 ND directions. 

\item[-] A D1-brane extended along $\xi$ $\Longrightarrow$ 6 ND directions. 

\item[-] An Euclidean D5-brane wrapped on ${\bf T}^6$ $\Longrightarrow$ 6 ND directions.

\end{itemize}
Additionally, a D3-brane is mutually BPS with an NS5-brane if the system has $4k$ ND directions. So for instance we can have:

\begin{itemize}

\item[-] A D3-brane wrapping  $(0,1)_i(0,1)_j(1,0)_k$ and point-like along $\xi$  $\Longrightarrow$ 4 or 8 ND directions. 

\item[-] An Euclidean D3-brane on  $(0,1)_i(1,0)_j(1,0)_k$ and along $\xi$  $\Longrightarrow$ 4 or 8 ND directions. 

\end{itemize}
All these objects are mutually BPS with the NS5-brane sources. If their orientation is reversed, they will still be BPS in the NS5-brane background, but preserving a disjoint set of supercharges. 

Upon three T-dualities along $\{x^1, x^2, x^3\}$, these D-branes are mapped to some of the type IIA D-branes discussed in section \ref{s:simple}. More precisely,  the D1-brane becomes a D4-brane wrapped on $\Pi_3^{\rm tor}$ and the Euclidean D5-brane is mapped to the Euclidean D2-brane wrapping the three-chain $\Sigma_3$. Similarly, the D3-branes become the D4-branes on the four-chains $\Sigma_4^i$ and the Euclidean D3-branes become the Euclidean D2-branes wrapped on $\Pi_2^i$. The configurations of AB strings and particles ending on monopoles and instantons described in section \ref{s:simple} have a clear microscopic origin in this dual type IIB setup. For instance, a D3-brane wrapped on $(0,1)_1(0,1)_2(1,0)_3$ intersects the NS5-brane at a point in ${\bf T}^6$ and is pointlike in 4d, by a Hanany-Witten effect (see e.g. \cite[Appendix B]{Berasaluce-Gonzalez:2012awn}) it must have $M$ D1-branes stretched along $\xi$ between the NS5-branes and its 4d location. The same occurs in the smeared description \eqref{Hex}, where the Freed-Witten anomaly induced by the $H$-flux on such a D3-brane is cured by the same stack of $M$ D1-branes.

\section{Massive $p$-from spectra in twisted tori}
\label{ap:spectra}

The aim of this appendix is to provide the explicit $p$-form eigenforms and eigenvalues of the twisted tori $\tilde{\bf T}^3$, whose $p$-form  spectrum was discussed in section \ref{s:direct} based on the method of  \cite{Ben_Achour_2016}. Furthermore, we compare the results with the analyses carried out in \cite{Andriot:2018tmb}, to show that the said spectrum is complete.

Consider the three-dimensional twisted torus  $\tilde{\bf T}^3$ with metric 
\be
ds^2_{\tilde{\bf T}^3} = (2\pi)^2 \left[(R_i\eta^i)^2 + (R_{j+3}\eta^{j+3})^2 + (R_{k+3}\eta^{k+3})^2 \right] \,,
\ee
and twist $d\eta^{i} = - N \eta^{j+3}\wedge \eta^{k+3}$. We choose angular coordinates $x^a\in[0,1]$ such that the $\eta^a$ are parameterized as 
\be
\eta^{i} = dx^{i} + N x^{k+3} dx^{j+3},\hspace{1em}\eta^{j+3} = dx^{j+3},\hspace{1em}\eta^{{k+3}}=dx^{{k+3}},
\ee
with $N=2\pi\lambda_{\chi} R_{k+3} R_{j+3}/R_{i}$ and  $\lambda_{\chi}\in \mathbb{R}$.

Then, we take the one-form Killing vector $\chi=2\pi R_{i} \eta^{i}/\sqrt{V}$, which satisfies the assumed properties presented in section \ref{s:direct}:
\begin{equation}
	\star d \chi = \lambda_\chi \,\chi \,, \qquad \Delta_3 \chi = \lambda_\chi^2\, \chi \,, \qquad \chi^2 = 1  \,, \qquad  \lambda_\chi, \in \IR \, ,
\end{equation}
where $V=8\pi^3R_iR_{j+3}R_{k+3}$ is the volume of $\tilde{\bf T}^3$.

Let $\{\phi_i\}$ represent the basis of complex scalar eigenforms of the Laplacian. The explicit shape of these eigenforms is \cite{Andriot:2018tmb}

\bea
&\phi_{p,q} = \frac{e^{2\pi i q x^{j+3}}e^{2\pi i p x^{k+3}}}{\sqrt{V}},\label{eq: phipq}\\
& \phi_{k,l,n} = \sqrt{\frac{2\pi R_{j+3}}{|N|V}}\frac{1}{\sqrt{2^n n!\sqrt{\pi}}} e^{2\pi i k(x^{i} + N x^{k+3} x^{j+3})}e^{2\pi l x^{k+3}} \sum_{m\in\mathbb{Z}}e^{2\pi i k mx^{k+3}}\Phi_n^\lambda(\omega_m),\label{eq: phikln}
\eea
with eigenvalues
\bea
& \sigma^2_{p,q}= \frac{q^2}{R^2_{j+3}}+\frac{p^2}{R^2_{k+3}}  ,\\
&\sigma^2_{k,l,n} = \frac{k^2}{R_{i}^2} + (2 n +1)k \,\frac{|\lambda_{\chi}|}{R_{i}},
\eea
where we have defined $\lambda=k\lambda_\chi /R_{i}$, $\omega_m= 2\pi R_{j+3}(x^{j+3} + \frac{m}{N}+\frac{l}{kN})$ and $\Phi_{n}^{\lambda}(z)=|\lambda|^{1/4}\Phi_n(|\lambda|^{1/2}z)$ with $\Phi_n$ being the Hermite polynomial. The integers have the following ranges
\be
p,q \in \mathbb{Z}^2,\,\,\,k\in \mathbb{Z}/\{0\},\,\,\, n\in \mathbb{N},\,\, \,l=0,1,\dots,|k|-1.
\ee

Moreover, the exact one-forms are given by the complete set $\{d\phi_i\}$. Based on the approach presented in section \ref{s:direct}, we observe that the sets of co-exact one-forms, denoted by $S_i$ and $T_i$ and defined in equation \eqref{eq: BandC}, exhibit closure properties when subjected to the $\star d$ operator. \newline 
Substituting the scalar eigenforms \eqref{eq: phipq} and \eqref{eq: phikln} into the sets $S_i$ and $T_i$, we obtain the eigenforms $U_{i}^{\pm}$ defined in \eqref{eq: Dpm}. Also, the corresponding eigenvalues $\lambda^{\pm}_i$ can be obtained from \eqref{eq: lambdapm}.

Using the first tower of scalar eigenforms, $\phi_{p,q}$, we acquire the following results:
\vspace{1em}

$\text{For}\,\, p,q \in \mathbb{Z}^2/\{0,0\}:$
\bea
 &U^{\pm}_{p,q} = \frac{2\pi\,\phi_{p,q}}{\sqrt{(\lambda^{\pm}_{p,q})^2 +\sigma_{p,q}^2 }} \left(   \frac{(\lambda^{\pm}_{p,q})^2-\sigma_{p,q}^2 }{i \lambda_{\chi}}\,\,R_{i} \,\eta^{i}- \frac{R_{j+3} }{R_{k+3}}\,p\,\eta^{j+3}+\frac{R_{k+3}}{R_{j+3}}\,q\,\eta^{k+3}\right),\nonumber\\
&(\lambda^\pm_{p,q})^2 = \frac{q^2}{R^2_{j+3}}+\frac{p^2}{R^2_{k+3}}  + \frac{ \lambda_\chi^2}{2} \pm \frac{ \lambda_\chi}{2}\sqrt{ \lambda_\chi^2 +\frac{4\,q^2}{R^2_{j+3}}+\frac{4\,p^2}{R^2_{k+3}}   } \,. 
\eea

For $p=q=0$:
\bea
& U^{-}=0, \hspace{1em} U^{+}_{0,0} = \frac{2\pi R_{i}}{\sqrt{V}}\,\eta^{i},\\
&\lambda^{+}_{0,0} = \lambda_{\chi}^2.
\eea

Comparing with the analysis carried out in \cite{Andriot:2018tmb} one can see that the above results reproduce exactly half of the co-exact one-form spectrum (see \cite{Andriot:2018tmb}, eq. (2.34) and Table 1).

Finally, for the second type of eigen-scalars, $\phi_{k,l,n}$, due to its very intricate expression (see eq. (2.59) from \cite{Andriot:2018tmb} for details), let us focus exclusively on the resulting eigenvalues and their degeneracy. Substituting again in \eqref{eq: lambdapm} we obtain 

\be
 (\lambda_{k,l,n}^{\pm})^2 = \frac{k^2}{R_{i}^2} + \frac{(2n +1)k |\lambda_{\chi}|}{R_{i}} + \frac{\lambda_\chi^2}{2} \pm \frac{\lambda_\chi}{2}\sqrt{\lambda_\chi^2 + \frac{4k^2}{R_{i}^2} + \frac{4(2n +1)k |\lambda_{\chi}|}{R_{i}}},\,\,\, n\in\mathbb{N}^*. \label{eq: lambdakln}
\ee

Therefore, we observe an exact correspondence with the remaining spectrum. It is important to note that we have shifted $n$ to $n+1$ in \eqref{eq: lambdakln}. This is because the solution \eqref{eq: lambdakln} with $n=0$ is in fact trivial using our method, and the matching with \cite{Andriot:2018tmb} occurs at our $n\geq1$.

%\newpage
%\addcontentsline{toc}{section}{References}
\bibliographystyle{JHEP2015}
\bibliography{biblio}

\end{document}